\documentclass{aa}
\usepackage{txfonts}
\usepackage{natbib}
\usepackage{graphicx}

\bibpunct[, ]{(}{)}{;}{a}{}{,}

\def\llm{{\sc LLmodels}}

\def\ddafit{{\sc DDAFit}}
\def\width{{\sc WIDTH9}}
\def\synthmag{{\sc SynthMag}}
\def\logg{\log g}
\def\teff{T_{\rm eff}}

\def\kms{km\,s$^{-1}$}

\def\bs{$\langle B \rangle$}
\def\paper1{Paper~I}
\def\halpha{H$\alpha$}
\def\hbeta{H$\beta$}

\def\ei{$E_{\rm i}$}
\def\loggf{$\log (gf)$}
\def\vt{$\xi_{\rm t}$}
\def\vsini{$v_{\rm e}\sin i$}
\def\eqw{$W_\lambda$}
\def\hd{$\alpha$\,Cir}

\newcommand{\txtbf}[1]{#1}

\defcitealias{Bruntt08}{BNC08}

\begin{document}

\title{A self-consistent empirical model atmosphere, abundance and stratification analysis of
the benchmark roAp star $\alpha$~Circini%
\thanks{Based on observations collected at the European Southern Observatory,
Paranal, Chile (ESO program 68.D-0254)}}
\titlerunning{A self-consistent empirical analysis of the benchmark roAp star $\alpha$~Circini}


\author{O. Kochukhov\inst{1} \and D. Shulyak\inst{2} \and T. Ryabchikova\inst{2,3}}
\offprints{O. Kochukhov, \email{oleg@fysast.uu.se}}
\institute{
Department of Physics and Astronomy, Uppsala University, Box 515, 751 20, Uppsala, Sweden \and
Institute of Astronomy, Vienna University, Turkenschanzstrasse 17, 1180 Vienna, Austria \and
Institute of Astronomy, Russian Academy of Science, Pyatnitskaya 48, 119017 Moscow, Russia
}

\date{Received 13 January 2009 / Accepted 15 March 2009}

\abstract
{Chemically peculiar (CP) stars are unique natural laboratories for investigation of the microscopic diffusion 
processes of chemical elements. The element segregation under the influence of gravity and radiation pressure 
leads to the appearance of strong abundance gradients in the atmospheres of CP stars. 
Consequently, the atmospheric temperature-pressure structure of these objects 
could deviate significantly from the atmospheres of normal stars with homogeneous abundances.}
{In this study we performed a self-consistent, empirical model atmosphere study of the brightest 
rapidly oscillating Ap star \hd. We account for chemical stratification in the model atmosphere calculations
and assess the importance of non-uniformed
vertical element distribution on the model structure, energy distribution and hydrogen line profiles.}
{For chemical stratification analysis we use the \ddafit\ minimization tool in combination with a magnetic
spectrum synthesis code. The model atmospheres with inhomogeneous vertical distributions of elements
are calculated with the \llm\ stellar model atmosphere code.}
{Based on an iterative procedure of the chemical abundance analysis of 52 ions of 35 elements,
stratification modeling of 4 elements \txtbf{(Si, Ca, Cr and Fe)} and subsequent re-calculations of
the atmospheric structure, we derived a new model atmosphere of \hd, which is consistent with the inferred
atmospheric chemistry of the star. We find $\teff$\,=\,7500~K, $\logg$\,=\,4.1, and 
demonstrate that chemical stratification has a noticeable impact on the model
structure and modifies the formation of the hydrogen Balmer lines.
At the same time, energy distribution appears to be less sensitive to the presence of large abundance
gradients.}
{Our spectroscopically determined $\teff$ of \hd\ agrees with the fundamental effective temperature of this
star. This shows that temperatures inferred in detailed spectroscopic analyses of cool magnetic CP stars
are not affected by a large systematic bias.}


\maketitle

\section{Introduction}

Atmospheres of \txtbf{certain} chemically peculiar (CP) B--F type stars are dynamically stable against
strong convective and turbulent motions. In combination with strong, large-scale
magnetic fields and slow rotation, this provides an ideal condition for the operation
of the microscopic particle diffusion \citep{michaud}. The overall, conspicuously non-solar, chemical signature
of the CP-star atmospheres is determined by an interplay between the effects of the gravitational
settling and radiative levitation of chemical species. Due to rapid variation of thermodynamic
properties in the outer stellar envelope, radiative diffusion is also able to produce a
vertical element separation within the line-forming atmospheric region of \txtbf{certain} CP stars.
The resulting picture of \textit{chemical stratification} 
depends upon the effective temperature and magnetic field strength of the star
\citep{TR08}. However, the light and iron-peak elements generally follow a simple step-like
behaviour with a large (relative to solar) 
element underabundance in the upper atmosphere
and an overabundance in the lower photospheric layers \cite[see, for example,][]{BWD01,RPK02,RLK05,vip}.
A qualitatively similar chemical stratification is predicted by a few available theoretical diffusion
calculations \citep{babel,LM04,LMH09}.
At the same time, rare-earth elements (REE) show completely opposite picture
with a large overabundance occurring in the superficial atmospheric layers and normal concentration or
underabundance in the continuum forming regions \citep{nd-NLTE,pr-NLTE}.

Due to large abundance anomalies and chemical gradients, the atmospheric structure of
CP stars can deviate significantly from that of the normal B--F stars with similar fundamental parameters and homogeneous
atmospheres. For example, the presence of core-wing anomaly in the Balmer line profiles of cool Ap stars indicates
that their temperature structure is \txtbf{most likely} different from normal stars \citep{cowley_cwa,kochukhov_cwa}. If unaccounted, these effects
may lead to significant systematic errors in determination of the stellar
atmospheric and fundamental parameters. 

In spite of recent advances
in the theoretical treatment of chemical stratification in stellar atmospheres \citep{diff_hui,LM04,diff_alecian,LMH09},
existing models cannot be applied for the routine analysis of CP-star atmospheres due to their complexity, uncertainties
in the description of relevant physical processes, incomplete atomic data, and occasional lack of self-consistency
between the model atmosphere structure and vertical abundance profiles.
Instead, empirical stratification studies, which derive element distributions from the observed line profiles,
typically employ model atmospheres computed with
the assumption of a homogeneous element distribution. 
This leads to an inconsistency between the inferred stratification 
and the adopted atmospheric temperature-pressure structure.

The aim of the present paper is to explore a new, alternative approach to the problem of
understanding the interplay between non-solar, inhomogeneous chemistry and the atmospheric
structure of CP stars. Similar to previous empirical stratification studies,  we use high-quality
spectroscopic observations to obtain the average chemical composition, vertical distribution \txtbf{of some of the most}
important elements, and place constraints on the stellar parameters. Then we make a second step of
including individual abundances as well as chemical stratification in the model atmosphere
calculation with our \llm\ code \citep{llm}. This two-stage procedure is repeated until convergence is achieved, providing us with a
self-consistent model atmosphere and reliable stellar parameters inferred from best available
observations.

We selected the bright ($V=3.2$), cool magnetic Ap star $\alpha$~Circini (HD~128898, HR~5463,
HIP~71908) for our investigation. This star is the brightest of about 40 currently known rapidly
oscillating Ap (roAp) stars and is one of the most promising objects for detailed asteroseismic
investigation. It pulsates in several non-radial oscillation modes \citep{kurtz} and was recently
observed photometrically by the WIRE satellite \citep{wire}. The time-resolved spectroscopic
observations of pulsations in  \hd\ \citep{RSK07,KRW07} allowed to probe the vertical mode
cross-section and study the dynamics of the outer atmospheric layers enriched in REEs. These studies
infer a very different pulsation behaviour of the lines of iron-peak and rare-earth elements, which
yields an independent confirmation of the validity of the chemical stratification picture outlined
above. The ongoing seismic studies of \hd\ require reliable fundamental parameters and detailed
knowledge of the stellar atmospheric structure.

A model atmosphere study of \hd\ may benefit from a rich collection of the photometric,
spectrophotometric, and high-resolution spectroscopic data collected for this star over several
decades. In particular, the availability of the observed UV and optical spectral energy 
distribution calibrated in absolute units is very valuable for testing theoretical model with
chemical stratification included. The absence of such data leaves significant uncertainty
in the parameter determinations of magnetic CP stars
\citep[see, for example,][]{hd137509,hd24712}.
Furthermore, the angular diameter of \hd\ was recently measured interferometrically
by \citet[][hereafter BNC08]{Bruntt08}. This allowed direct determination of the stellar radius for the first time for
any magnetic Ap star. The effective temperature could be subsequently inferred
from the bolometric flux,
almost independently from the model atmosphere techniques (however, still
using theoretically calculated flux for the IR region 
due to the lack of spectrophotometric observations). Thus, \hd\ is a unique benchmark cool Ap star
and an extremely useful object for the verification and calibration of various photometric and spectroscopic
techniques of deriving basic
stellar properties.  
In fact,
the $\teff$\,=\,$7420\pm170$~K found by \citetalias{Bruntt08} is lower than \textit{all} previous photometric and
spectroscopic temperature estimates, which span the range of 600~K. This new fundamental temperature
of \hd\ differs significantly from  $\teff$\,=\,$7900\pm200$~K suggested for this star in the
earlier model atmosphere analysis \citep{KRW96}, raising doubts about the 150--200~K $\teff$
accuracy often claimed in the studies of cool Ap stars.

The chemical composition of the atmosphere of \hd\ was first studied by \citet{KRW96}. Based on new
fundamental values of the atmospheric parameters, \citetalias{Bruntt08} redetermined abundances for a
number of chemical elements using models with solar metallicity and with individual abundances. However,
they did not carry out a stratification analysis but assumed a chemically-homogeneous  atmosphere
both for the line formation and for model atmosphere calculation. On the other hand, \citet{RKB08} presented
an analysis of the Ca stratification in \hd\ based on the model parameters from \citet{KRW96} and 
neglecting the influence of chemical gradients on the model structure.

In this paper we determine abundances and reconstruct the stratification of most important elements
in the atmosphere of \hd. We then explore the effect of this non-solar, inhomogeneous atmospheric
chemistry on the derivation of stellar parameters. Since the non-uniform element distribution
changes the model temperature-pressure structure, we implement an iterative procedure of the
stratification and abundance analysis with the subsequent re-calculation of the model atmosphere
and re-determination of stellar parameters.

This paper is organized as follows.
In the next section we give an overview of observational data.  Then, in Sect.~\ref{sec:methods} we
describe the general methods of chemical abundance and stratification analysis, 
model atmosphere calculation and atomic line data used.  In Sect.~\ref{sec:results} the results of the stratification
modeling are presented and the influence of chemical gradients on various observed stellar
characteristics is discussed. The summary of our investigation and discussion of our results
in the context of other recent studies of cool Ap stars are presented in
Sect.~\ref{sec:concl}.

\section{Observations}
\label{sec:obs}

For abundance and stratification analysis we used the spectrum of \hd\ obtained with the UVES
instrument \citep{uves} at the ESO VLT. The observations were carried out in two dichroic modes,
using a 0.5\arcsec\ slit for a spectral resolution of 80\,000. The spectra cover the full wavelength
region of 3040--10\,400~\AA\ except for a few small gaps. Reduction of the UVES spectra was performed
with the automatic pipeline \citep{pipeline}. The reduction process was described in more detail by
\citet{vip} and \citet{RKB08}.

The observed ultraviolet spectral energy distribution of \hd\ was obtained from IUE Newly Extracted
Spectra (INES) archive\footnote{{\tt http://sdc.laeff.inta.es/ines/index2.html}}. We used the
average of rebinned, high-dispersion spectra, recorded with a large aperture in the 1150--3100~\AA\
wavelength window. For the spectral interval 3200--7500~\AA\ we used low-resolution energy
distributions of \hd\ from the catalogues by \citet{burnashev} and \citet{alekseeva}. These data are
calibrated in absolute flux and show a systematic difference of 5--10\% below $\lambda=4500$~\AA.
The agreement is better than 5\% for the longer wavelengths.

The spectrophotometry of \hd\ was complemented with the broad-band photometric measurements in the
UBVRI \citep{ubvri} and JHKLM \citep{jhklm} systems. Stellar magnitudes were converted to absolute
flux units with the help of calibrations by \citet{bessell} and \citet{engels}, respectively.

In addition to using the low-resolution spectral energy distribution as well as metal and hydrogen line profiles in
the UVES spectra, we compared predictions of our theoretical models with the Str\"omgren and Geneva photometric
parameters of \hd. The observed Geneva colors were obtained from the catalogue by \citet{geneva}. For the 
Str\"omgren photometry, numerous measurements listed for \hd\ in the SIMBAD database\footnote{{\tt
http://simbad.u-strasbg.fr/simbad/}} yield average parameters $b-y$\,=\,$0.135\pm0.013$, $m_1$\,=\,$0.218\pm0.020$,
$c_1$\,=\,$0.783\pm0.006$, and H$\beta$\,=\,$2.829\pm0.009$.

\section{Methods}
\label{sec:methods}

\subsection{Abundance and stratification analysis}

The classical abundance analysis in the approximation of chemically-homogeneous atmosphere was carried out with the
help of the equivalent width technique. We used the \width\ code \citep{a9}, modified by V.~Tsymbal
\citep[see][]{RPK02}, to estimate abundances of 35 elements for a given model atmosphere. In few cases of blended
lines or lines affected by the hyperfine structure (hfs) theoretical line profiles calculations were performed with
the magnetic spectrum synthesis code \synthmag\ \citep{synthmag07}, assuming a homogeneous radial field structure 
with the magnetic
field modulus of \bs\,=\,2.0\,kG. This value corresponds to the upper limit of the  field strength in \hd\ estimated by
\citet{KRW96} and \citetalias{Bruntt08}. We also tried \bs\,=\,1.0\,kG and found no significant effect on the results of
abundance or stratification analysis.

Since it is not possible to account for the Zeeman splitting and polarized radiative transfer in the \width\ code, we
used a pseudo-microturbulence to mimic the magnetic intensification effects \citep[see][]{KRW96}. This approach is
appropriate for the equivalent width analysis of magnetic stars with \bs\,$\le$\,3\,kG. The pseudo-microturbulence
parameter of \vt\,=\,$1.2\pm0.2$~\kms\ was estimated by matching profiles of the \ion{Fe}{i} and \ion{Fe}{ii} lines
computed with the full treatment of polarized radiative transfer and zero microturbulence by a non-magnetic spectrum
synthesis calculation. Since \hd\ has an inhomogeneous surface abundance structure (\citealt{KR01}; \citetalias{Bruntt08}) and a
non-uniform surface magnetic field distribution, results of the abundance analysis and the formal \vt\ value may slightly
change with  the rotation phase. However, the spectral and magnetic variation of this star over its 4.46$^{\rm d}$
rotation period are minor compared to other cool Ap stars \citep[e.g.,][]{hr3831}, which excludes a large influence
of the horizontal inhomogeneities on the results of our analysis.

Stratification modeling was performed for Si, Ca, Cr, and Fe. These elements have sufficient number of lines in the
observed wavelength region and are able to influence the model structure and flux distribution by altering the line
blanketing. For these four stratified elements we adopted the step-function approximation of the vertical abundance 
distribution \citep{RLK05,RKB08}. The vertical abundance profile was described by four parameters: chemical abundance
in the upper atmosphere, abundance in deep layers, the position of the abundance jump and the width of the transition
region where chemical abundance changes between the two values. All four parameters were optimized simultaneously with
the IDL-based \ddafit\ least-squares fitting procedure \citep{synthmag07}. This code adjusts the vertical abundance
distribution of a given element to achieve the best fit to a set of observed spectral line profiles. 
In these calculations we used the \synthmag\ code and adopted \vsini\,=\,12.5~\kms.

We note that the assumption of constant abundances in the upper atmosphere and in the lower photospheric
layers  is related to the lack of spectral lines sensitive to the abundance changes in these atmospheric
regions. Due to substantial rotation of \hd\ our choice of lines for the stratification analysis is limited to
absorption features probing a certain line-forming region, which does not extend over the entire atmosphere. In
particular, we cannot probe layers with $\log\tau_{5000}\gg 0$. The cores of strong lines could potentially
provide us with the information about stratification in the uppermost layers, but such lines are not available
for the majority of elements. We estimated that robust determination of stratification using the lines
available in the spectrum of \hd\ is possible for the $\log\tau_{5000}\approx[-3.0,+0.5]$ region. Outside this
range of optical depths, an unconstrained solution of the vertical inversion problem is non-unique due to
ill-conditioned nature of the problem \citep{vip}. Thus, to overcome this difficulty, we force chemical
abundance to a constant value in these layers.

\subsection{Atomic line data}

\begin{table*}[!th]
\caption{List of Si, Cr, and Fe spectral lines used for stratification 
calculations. Columns give the ion identification, central wavelength, excitation
potential, oscillator strength (\loggf) and the Stark damping constant
($\log\,\gamma_{\rm St}$) for $T=10\,000$\,K. The last column gives reference for the adopted oscillator strength.}
\label{tab:list}
\begin{footnotesize}
\begin{center}
\begin{tabular}{lcrrrl|lcrrrl}
\noalign{\smallskip}
\hline
\hline
Ion &$\lambda$ (\AA) &\ei\,(eV)  &\loggf&$\log\,\gamma_{\rm St}$ & Ref.&Ion &$\lambda$ (\AA) &\ei\,(eV)  &\loggf&$\log\,\gamma_{\rm St}$& Ref.\\
\hline
\ion{Si}{ii}&  5055.984& 10.074 & 0.593&  -4.78 &SG   &\ion{Cr}{ii}&  7419.646&	 4.750 &-2.64~&  -6.630&RU  \\  
\ion{Si}{ii}&  5056.317& 10.074 &-0.359&  -4.78 &SG   &            &          &        &      &        &    \\            
\ion{Si}{i} &  5690.425&  4.930 &-1.76~&  -4.57 &NIST &\ion{Fe}{i} &  4250.120&	 2.469 &-0.405&  -5.410&MFW \\           
\ion{Si}{i} &  5948.54 &  5.082 &-1.23~&  -4.45 &NIST &\ion{Fe}{i} &  4271.760&	 1.485 &-0.164&  -6.070&MFW \\        
\ion{Si}{ii}&  5957.559& 10.067 &-0.301&  -5.02 &SG   &\ion{Fe}{ii}&  4508.288&	 2.856 &-2.35~&  -6.530&RU  \\        
\ion{Si}{ii}&  5978.930& 10.074 & 0.004&  -5.01 &SG   &\ion{Fe}{ii}&  4576.340&	 2.844 &-2.98~&  -6.530&RU  \\         
\ion{Si}{i} &  6142.483&  5.619 &-1.420&  -3.57 &astr &\ion{Fe}{ii}&  4635.316&	 5.956 &-1.58~&  -6.560&RU  \\         
\ion{Si}{ii}&  6371.371&  8.121 &-0.030&  -5.32 &BBCB &\ion{Fe}{ii}&  4923.927&	 2.891 &-1.50~&  -6.500&RU  \\         
\ion{Si}{i} &  7003.569&  5.964 &-1.78~&  -4.28 &astr &\ion{Fe}{i} &  4924.770&	 2.279 &-2.241&  -5.890&BK  \\
\ion{Si}{i} &  7034.901&  5.871 &-1.780&  -3.63 &astr &\ion{Fe}{ii}&  4993.358&	 2.807 &-3.68~&  -6.530&RU  \\         
\ion{Si}{i} &  7944.001&  5.984 &-1.310&  -3.93 &NIST &\ion{Fe}{i} &  4994.130&	 0.915 &-3.080&  -6.150&MFW \\
\ion{Si}{i} &  8892.720&  5.984 &-0.830&  -4.37 &astr &\ion{Fe}{ii}&  5045.114&	10.308 & 0.00~&  -5.250&RU  \\         
            &          &        &      &        &     &\ion{Fe}{i} &  5049.820&	 2.279 &-1.355&  -5.910&BWLW\\         
\ion{Cr}{ii}&  3484.147&  2.455 &-2.14~&  -6.611&RU   &\ion{Fe}{ii}&  5061.718&	10.308 & 0.28~&  -5.330&RU  \\            
\ion{Cr}{i} &  4274.797&  0.000 &-0.231&  -6.240&MFW  &\ion{Fe}{ii}&  5149.465&	10.448 & 0.55~&  -5.180&RU  \\          
\ion{Cr}{ii}&  4634.070&  4.072 &-1.236&  -5.359&RU   &\ion{Fe}{i} &  5165.410&	 4.220 &-0.40~&  -4.710&astr\\         
\ion{Cr}{i} &  4652.157&  1.004 &-1.03~&  -6.297&MFW  &\ion{Fe}{i} &  5198.711&	 2.223 &-2.135&  -6.070&MFW \\         
\ion{Cr}{ii}&  4901.623&  6.487 &-1.14~&  -6.792&RU   &\ion{Fe}{ii}&  5291.666&	10.480 & 0.54~&  -5.260&RU  \\         
\ion{Cr}{i} &  4922.265&  3.104 &~0.27~&  -6.193&MFW  &\ion{Fe}{i} &  5365.399&	 3.573 &-1.02~&  -6.250&BWLW\\        
\ion{Cr}{i} &  4936.336&  3.113 &-0.34~&  -6.205&MFW  &\ion{Fe}{ii}&  5427.826&	 6.724 &-1.58~&  -6.500&RU  \\         
\ion{Cr}{ii}&  5116.049&  3.714 &-3.64~&  -6.584&RU   &\ion{Fe}{i} &  5434.523&	 1.011 &-2.122&  -6.220&MFW \\         
\ion{Cr}{ii}&  5174.850&  6.869 &-2.42~&  -6.792&RU   &\ion{Fe}{ii}&  5439.707&	 6.729 &-2.38~&  -6.500&RU  \\  
\ion{Cr}{i} &  5204.511&  0.941 &-0.208&  -6.154&MFW  &\ion{Fe}{i} &  5576.089&	 3.430 &-1.000&  -5.390&MFW \\  
\ion{Cr}{ii}&  5246.768&  3.714 &-2.56~&  -6.660&RU   &\ion{Fe}{ii}&  6084.111&	 3.199 &-3.880&  -6.530&RU  \\  
\ion{Cr}{i} &  5296.691&  0.983 &-1.400&  -6.120&MFW  &\ion{Fe}{i} &  6127.907&	 4.143 &-1.399&  -6.020&BWLW\\  
\ion{Cr}{i} &  5297.377&  2.900 & 0.167&  -3.807&MFW  &\ion{Fe}{i} &  6165.360&	 4.143 &-1.474&  -6.020&BWLW\\  
\ion{Cr}{i} &  5304.180&  3.464 &-0.692&  -5.301&MFW  &\ion{Fe}{i} &  6335.331&	 2.198 &-2.177&  -6.160&BWLW\\  
\ion{Cr}{ii}&  5310.687&  4.072 &-2.41~&  -6.643&RU   &\ion{Fe}{i} &  6336.824&	 3.686 &-0.856&  -5.380&BK  \\          	  
\ion{Cr}{ii}&  5502.067&  4.168 &-2.12~&  -6.630&RU   &\ion{Fe}{ii}&  6432.680&	 2.891 &-3.69~&  -6.500&RU  \\        			  
\ion{Cr}{ii}&  5564.741& 10.893 & 0.51~&  -5.364&RU   &\ion{Fe}{i} &  6820.372&	 4.638 &-1.320&  -4.470&MFW \\        		
\ion{Cr}{ii}&  5678.390&  6.484 &-1.50~&  -6.627&RU   &\ion{Fe}{i} &  7155.630&	 5.010 &-0.724&  -4.820&K   \\         
\ion{Cr}{ii}&  6112.261&  4.745 &-2.98~&  -6.656&RU   &\ion{Fe}{i} &  7221.202&	 4.559 &-1.470&  -5.110&K   \\         
\ion{Cr}{i} &  6661.075&  4.193 &-0.19~&  -5.002&MFW  &\ion{Fe}{ii}&  7222.394&	 3.889 &-3.40~&  -6.520&RU  \\           
\ion{Cr}{i} &  7400.249&  2.900 &-0.111&  -5.326&MFW  &            &          &        &      &        &    \\  
\hline                                
\end{tabular}
\end{center}
SG -- \citet{Schulz};  NIST -- \citet{NIST}; BBCB -- \citet{berry};  
RU -- \citet[][ftp://ftp.wins.uva.nl/pub/orth]{Raassen1998}; K -- \citet{ATOMS}; 
MWF -- \citet{Martin}; BK -- \citet{bard}.
\end{footnotesize}
\end{table*}

Atomic data for the abundance calculations were extracted from the recently updated VALD database  \citep[][see also
http://vald.inasan.ru/$\sim$vald/ and references therein]{vald1,vald2}. For \ion{Tb}{iii} and \ion{Dy}{iii}
transition probabilities were kindly provided by A.~Ryabtsev (private communication). These calculations are based on
the extended analysis of the energy level structure.  For several ions (\ion{Mn}{i,ii}, \ion{Co}{i}, \ion{Eu}{ii},
\ion{Ho}{ii})  with substantial hfs splitting we included this effect in the abundance calculations.
The corresponding hfs constants were taken from \citet{ATOMS} for \ion{Mn}{i},
from \citet{HSR99} for \ion{Mn}{ii}, from \citet{hfs-Co1} for \ion{Co}{i}, from 
\citet{Eu2} for \ion{Eu}{ii}, and from \citet{Ho2} for \ion{Ho}{ii}.

Stratification analysis of Si, Ca, Cr, and Fe requires accurate and homogeneous line parameters for these elements.
For Ca we used the same lines as in \citet{RKB08}.  The line list for other elements is given in
Table\,\ref{tab:list}. 
The transition probabilities were
improved by fitting the solar flux spectrum for a few \ion{Si}{i} lines and one \ion{Fe}{i} line.
For \ion{Si}{ii} lines the Stark constants are taken from \citet{Wilke03},  while for all
other lines they are taken from Kurucz database \citep{ATOMS}.

\subsection{Model atmosphere calculations}

Our model atmosphere calculations are based on the version 8.6 of the 1-D, LTE model atmosphere code \llm\
\citep{llm}. This program treats the bound-bound opacity by direct, line-by-line spectrum synthesis and
is able to account for the effects of individual non-solar abundance and inhomogeneous vertical distribution
of elements. Chemical abundances and stratification are provided as input parameters for the \llm\ code 
and kept constant in the model atmosphere calculation process.

For all model atmospheres calculations in this paper we used a 72-layer vertical grid with equidistant spacing in the
$\log \tau_{5000}$ range between $-7$ and 2. The frequency-dependent quantities were  calculated between 50~\AA\ and
100\,000~\AA\ with a constant wavelength step of 0.1~\AA. The VALD database 
was used as the main
source of the atomic data for computation of the line absorption coefficients. As it was shown by
\citet{zeeman_paper1} and \citet{zeeman_paper2},  magnetic fields with a strength of 1--3~kG expected for \hd\ could be safely
neglected  in the model atmosphere calculations. Consequently, we ignored the effects of Zeeman splitting  and
polarized radiative transfer on the atmospheric structure.  However, we used a pseudo-microturbulent velocity of
1~\kms\ to qualitatively treat magnetic intensification and for consistency with the abundance analysis.

The studies by \citet{pr-NLTE,nd-NLTE} demonstrated that the line formation of rare-earth elements Pr and Nd can
strongly deviate from the local thermodynamic equilibrium. These elements are overabundant in the atmospheres of CP
stars by several dex compared to the solar atmosphere and have rich spectra dominated by the lines of the first and
second ions. The doubly ionized lines of Pr, Nd, and of some other REEs are unusually strong due to combined effect of
stratification of these elements and departures from LTE. Therefore, a NLTE stratification treatment of REEs is
essential for detailed spectrum synthesis and also for realistic estimate of their contribution to the line opacity
in the model atmosphere calculations. However, a NLTE analysis is beyond the scope of this paper and currently could
not be coupled to a model calculation. The large \vsini\ of \hd\ compared to other cool Ap stars also makes it
difficult to model stratification of REEs since some useful weak lines are not accessible for measurement. Therefore,
we followed  \citet{hd24712} in using a simplified treatment of the REE absorption. One can approximately reproduce
the observed relative line strengths of the Pr, Nd, Dy, and Tb ions by using abundance derived from the second ions and
reducing the oscillator strengths for the singly ionized REE lines. The adopted reduction factor of 1.5~dex is close
to the mean difference between concentrations of singly and doubly ionized REEs estimated in our classical abundance
analysis of \hd. With this scaling procedure the Pr, Nd, Dy, and Tb ionization equilibria are conserved in the model
atmosphere calculations while the line strength roughly matches the observed one for both ionization stages.

\subsection{Self-consistent atmospheric modeling}
\label{sec:ss}

Previous chemical stratification studies used high-resolution observations of CP stars to perform average abundance 
determination and to restore vertical abundance profiles for a few chemical elements. Generally, this analysis is
performed by interpreting observed profiles of spectral lines with different strength, excitation potential, and
ionization degree for a given element. However, the shapes and strengths of the line profiles predicted by theoretical
models may also indirectly depend on the vertical distribution of chemical elements through the influence of line
opacities, modified by peculiar abundances and chemical  stratification, on the temperature-pressure structure of the
model atmosphere. Thus, the stratification and abundance analyses are linked with the model atmosphere calculation.
The extent of the feedback of stratification and non-solar chemical composition on the model structure cannot be
estimated \textit{a priori}. Consequently, one has to apply an iterative procedure of stratification and abundance 
analysis followed by re-calculation of the model atmosphere structure. For the present analysis of \hd\ this 
procedure included the following main steps:
\begin{enumerate}
\item
calculation of a model atmosphere grid using mean abundances derived in previous studies;
\item
determination of $\teff$ and $\logg$ with the help of spectrophotometry, Balmer line profiles and broad-band colors;
\item
determination of the average abundances and chemical stratification using high-resolution spectra;
\item
calculation of a new model atmosphere grid, taking into account individual abundances and stratification 
found in the previous step;
\item
repeating the entire process starting from step~2 until convergence is achieved.
\end{enumerate}
The model atmosphere grids were calculated for the $\teff$ range of 7000--8000~K with a 200~K step
and for $\logg$\,=\,3.8--4.2 with a 0.1~dex step. We adopted the convergence criterium of the absence of significant
changes of the mean abundances, stratification profiles of studied chemical elements 
and model atmosphere parameters.

The final self-consistent, chemically-stratified model atmosphere is expected to reproduce
simultaneously the observed photometry, energy distribution, hydrogen line profiles and metallic line spectra. The
method of self-consistent analysis of a cool Ap-star atmosphere applied here is similar to the procedure described by
\citet{hd24712} for the roAp star HD\,24712. However, in practice robust re-determination of the atmospheric 
parameters is not possible for the latter star due to the lack of observational data. In contrast, \hd\ lends 
itself to a thorough study of the interplay between atmospheric chemistry and stellar parameters.

\begin{figure*}[!th]
\centerline{\includegraphics[width=15cm]{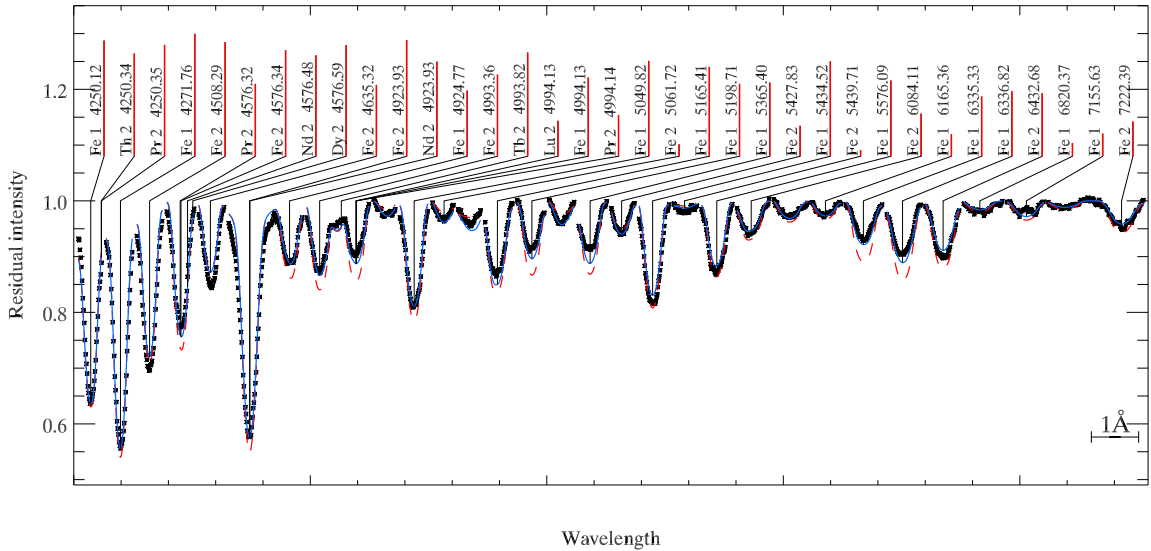}}
\caption{Comparison of the observed Fe line profiles (symbols) and calculations for the final
stratified abundance distribution (\textit{solid line}) and for the best-fit homogeneous abundance of Fe
(\textit{dashed line}).}
\label{fig:fe}
\end{figure*}

\onlfig{2}{
\begin{figure*}[!th]
\centerline{\includegraphics[width=15cm]{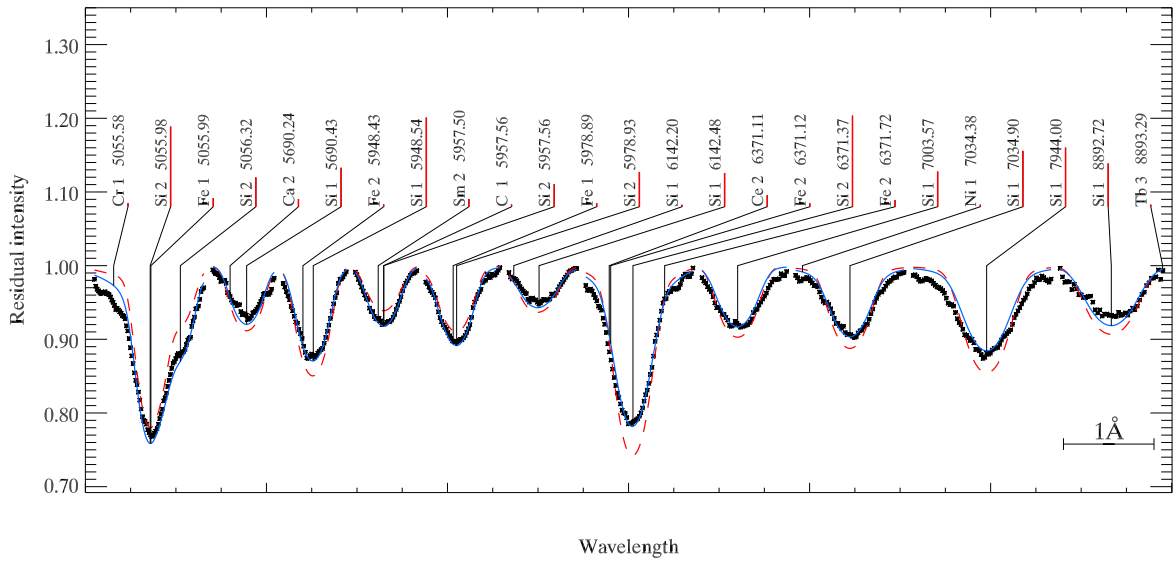}}
\caption{Same as in Fig.~\ref{fig:fe} but for Si.}
\label{fig:si}
\end{figure*}
}

\onlfig{3}{
\begin{figure*}[!th]
\centerline{\includegraphics[width=15cm]{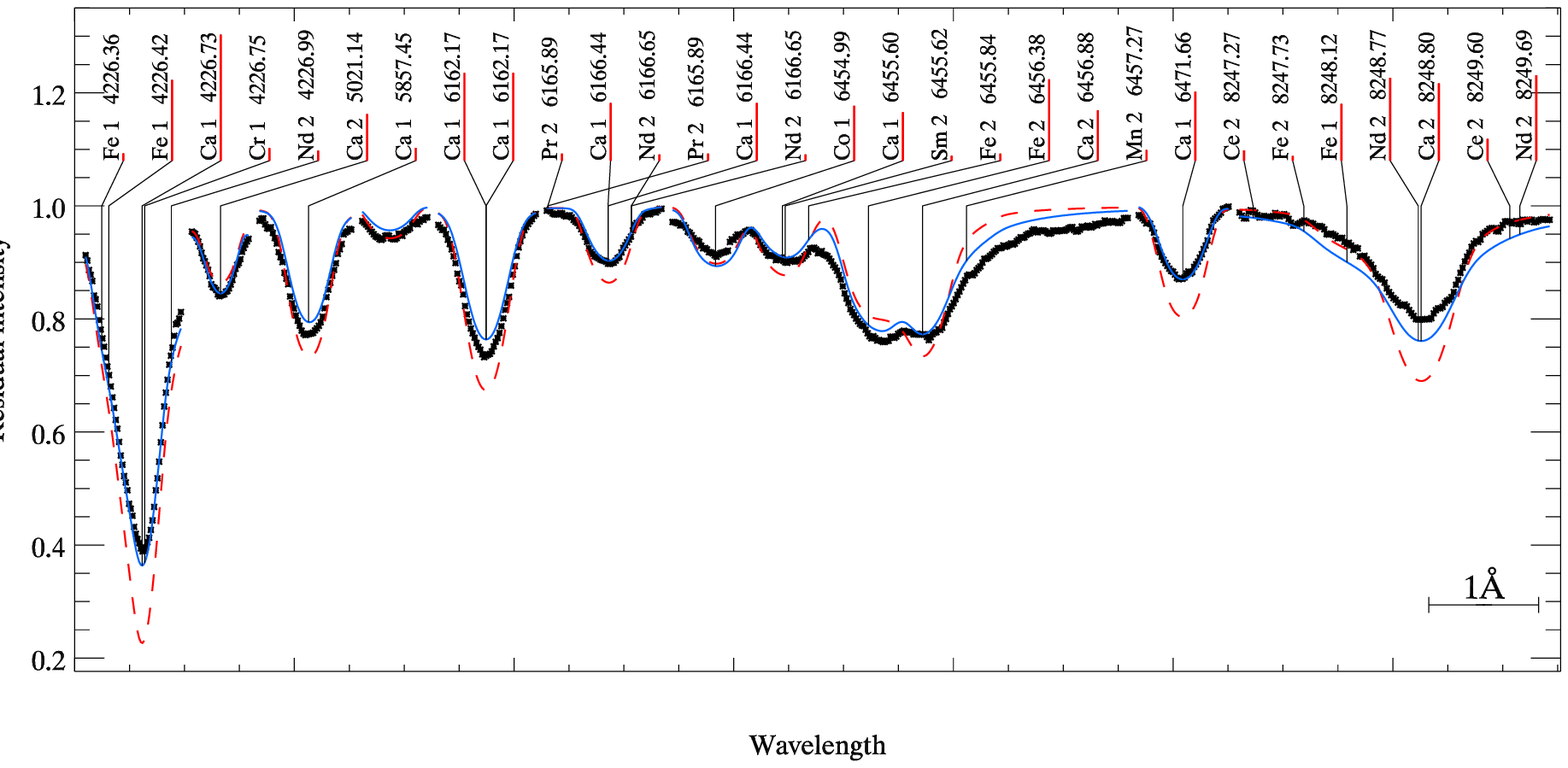}}
\caption{Same as in Fig.~\ref{fig:fe} but for Ca.}
\label{fig:ca}
\end{figure*}
}

\onlfig{4}{
\begin{figure*}[!th]
\centerline{\includegraphics[width=15cm]{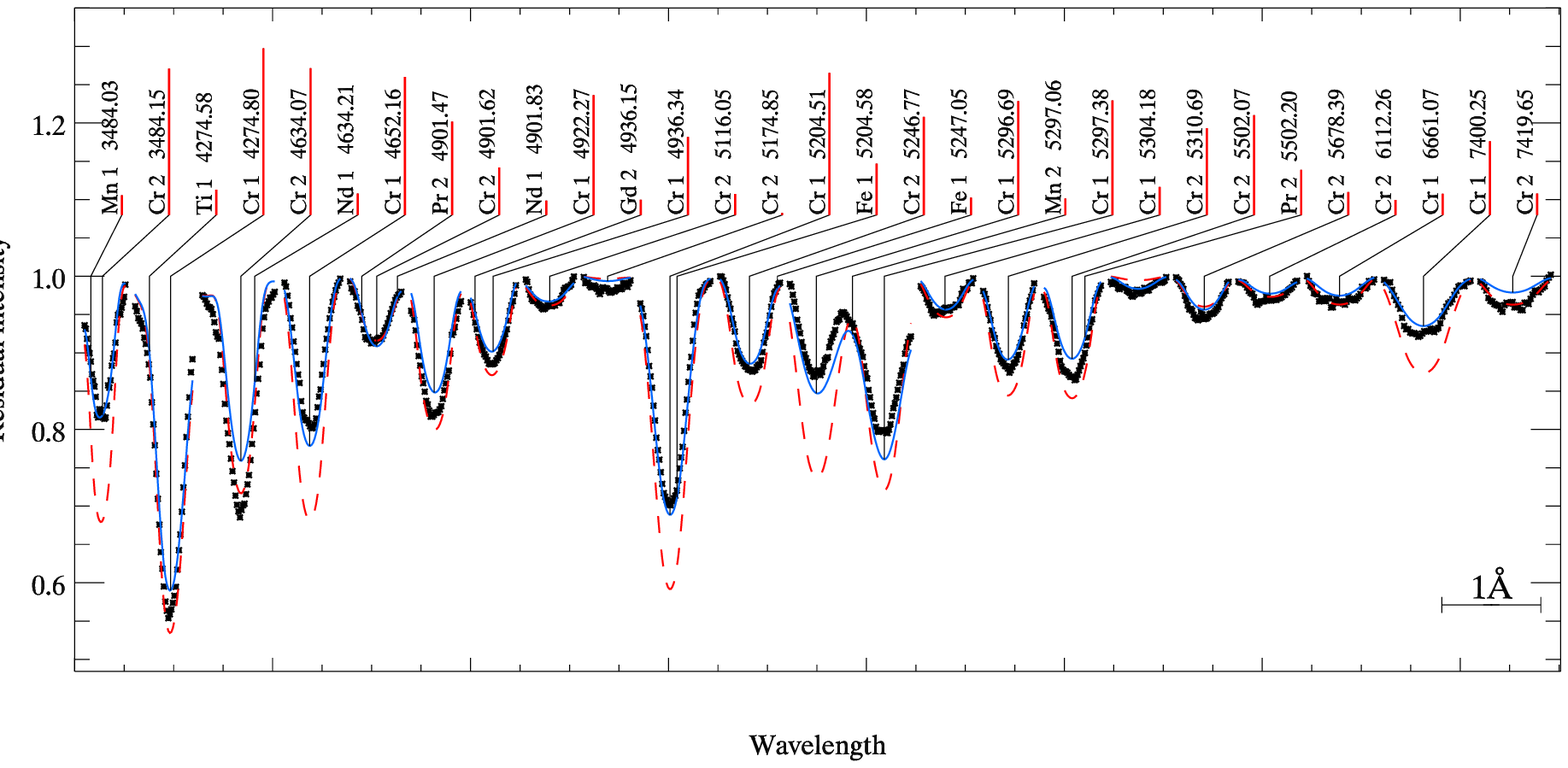}}
\caption{Same as in Fig.~\ref{fig:fe} but for Cr.}
\label{fig:cr}
\end{figure*}
}

\section{Results}
\label{sec:results}

\begin{figure}[!th]
\includegraphics[width=\hsize]{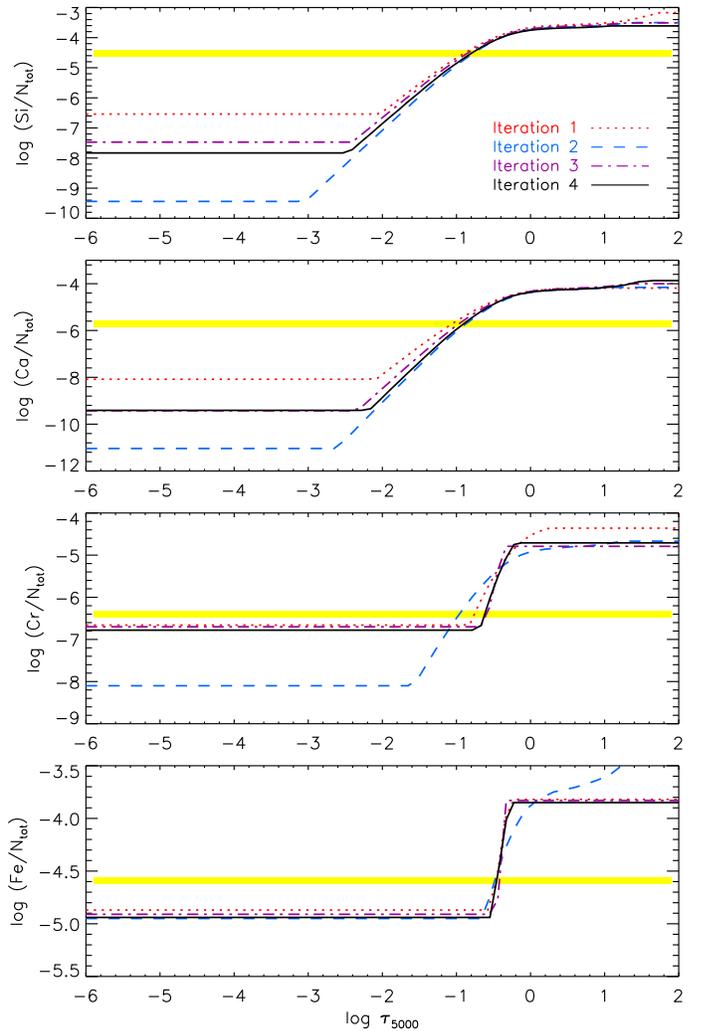}
\caption{Vertical distribution of Si, Ca, Cr, and Fe abundance in the atmosphere of \hd\ for different iterations
of the stratification analysis. The thick solid line corresponds to the final element distribution. The light
horizontal bar shows solar abundances. 
}
\label{fig:strat}
\end{figure}

\subsection{Chemical abundance pattern}
\begin{table*}[!th]
\caption{Abundances in the atmosphere of \hd\ with the error
estimates based on $n$ measured lines. Abundances derived in our study are compared with those reported by \citet{KRW96} 
and \citetalias{Bruntt08}. The solar abundances are from \citet{asplund}.}
\label{tab:abund}
\begin{footnotesize}
\begin{center}
\begin{tabular}{l|c|c|lr|lr|c}
\noalign{\smallskip}
\hline
\hline
Ion & \citet{KRW96} & \citetalias{Bruntt08}  & \multicolumn{4}{|c|}{This paper} &Sun \\                                  
    &$\log (El/N_{\rm tot})$ &$\log (El/N_{\rm tot})$ & $\log (El/N_{\rm tot})$ & $n$ & with HFS & $n$ &$\log (El/N_{\rm tot})$  \\       
\hline
\ion{C}{i}     & -4.00 & -3.60 & ~~$-$3.92$\pm$0.20 & 14 &                  &    &~~$-$3.65~ \\                        
\ion{N}{i}     & -4.40:&       & ~~$-$4.53$\pm$0.07 &  4 &                  &    &~~$-$4.26~ \\                        
\ion{O}{i}     & -3.80 & -3.53 & ~~$-$3.58$\pm$0.22 &  5 &                  &    &~~$-$3.38~ \\                        
\ion{Na}{i}    & -5.90 & -6.08 & ~~$-$5.91$\pm$0.06 &  2 &                  &    &~~$-$5.87~ \\                         
\ion{Mg}{i}    &       & -4.52 & ~~$-$4.52$\pm$0.15 &  7 &                  &    &~~$-$4.51~ \\                         
\ion{Mg}{ii}   & -4.46:&       & ~~$-$4.36:         &  1 &                  &    &~~$-$4.51~ \\                         
\ion{Al}{i}    &       & -5.49 & ~~$-$5.52:         &  1 &                  &    &~~$-$5.67~ \\                         
\ion{Si}{i}    & -4.20 & -4.67 & ~~$-$4.51$\pm$0.20 & 15 &                  &    &~~$-$4.53~ \\                         
\ion{Si}{ii}   &       &       & ~~$-$4.20$\pm$0.15 &  4 &                  &    &~~$-$4.53~ \\                         
\ion{S}{i}     &       & -4.55 & ~~$-$5.01$\pm$0.19 &  6 &                  &    &~~$-$4.90~ \\                           
\ion{K}{i}     &       &       & ~~$-$7.22:         &  1 &                  &    &~~$-$6.96~ \\                         
\ion{Ca}{i}    & -5.15 & -5.54 & ~~$-$5.29$\pm$0.16 & 13 &                  &    &~~$-$5.73~ \\                           
\ion{Ca}{ii}   &       &       & ~~$-$5.09$\pm$0.26 &  5 &                  &    &~~$-$5.73~ \\                           
\ion{Sc}{ii}   & -9.80 & -9.41 & ~~$-$9.39$\pm$0.19 &  7 &                  &    &~~$-$8.99~ \\                         
\ion{Ti}{i}    & -6.87 & -7.34 & ~~$-$7.33$\pm$0.19 &  3 &                  &    &~~$-$7.14~ \\                         
\ion{Ti}{ii}   & -6.87 & -7.31 & ~~$-$7.23$\pm$0.17 & 11 &                  &    &~~$-$7.14~ \\                         
\ion{V}{i}     & -7.65 & -7.44 & ~~$-$7.48:         &  1 &                  &    &~~$-$8.04~ \\                         
\ion{V}{ii}    &       &       & ~~$-$7.41$\pm$0.24 &  8 &                  &    &~~$-$8.04~ \\                         
\ion{Cr}{i}    & -5.55 & -5.92 & ~~$-$5.70$\pm$0.21 & 23 &                  &    &~~$-$6.40~ \\                         
\ion{Cr}{ii}   & -5.55 & -5.91 & ~~$-$5.46$\pm$0.35 & 21 &                  &    &~~$-$6.40~ \\                         
\ion{Mn}{i}    & -6.00 & -5.97 & ~~$-$6.14$\pm$0.20 & 19 &~~$-$6.41$\pm$0.21& 14 &~~$-$6.65~ \\                       
\ion{Mn}{ii}   & -6.00 & -5.97 & ~~$-$5.98$\pm$0.14 &  7 &~~$-$6.20$\pm$0.10&  7 &~~$-$6.65~ \\                       
\ion{Fe}{i}    & -4.50 & -4.71 & ~~$-$4.61$\pm$0.17 &116 &                  &    &~~$-$4.59~ \\                       
\ion{Fe}{ii}   & -4.50 & -4.62 & ~~$-$4.39$\pm$0.17 & 30 &                  &    &~~$-$4.59~ \\                       
\ion{Co}{i}    & -5.50 & -5.92 & ~~$-$5.93$\pm$0.26 & 27 &~~$-$6.35$\pm$0.36&  8 &~~$-$7.12~ \\                       
\ion{Co}{ii}   & -5.50 &       & ~~$-$5.63$\pm$0.31 &  5 &                  &    &~~$-$7.12~ \\                         
\ion{Ni}{i}    & -6.15 & -6.39 & ~~$-$6.51$\pm$0.26 &  8 &                  &    &~~$-$5.81~ \\                         
\ion{Se}{i}    &       &       & ~~$-$8.40:         &  1 &                  &    &~~$-$8.71~ \\                         
\ion{Sr}{i}    &       &       & ~~$-$7.27$\pm$0.13 &  5 &                  &    &~~$-$9.12~ \\                         
\ion{Sr}{ii}   & -7.25 &       & ~~$-$8.70$\pm$0.50 &  2 &                  &    &~~$-$9.12~ \\                         
\ion{Y}{i}     &       &       & ~~$-$7.90:         &  1 &                  &    &~~$-$9.83~ \\                         
\ion{Y}{ii}    & -8.50 & -8.95 & ~~$-$8.65$\pm$0.17 & 11 &                  &    &~~$-$9.83~ \\                         
\ion{Zr}{ii}   & -9.00 & -9.08 & ~~$-$9.34$\pm$0.12 &  3 &                  &    &~~$-$9.45~ \\                         
\ion{Ba}{ii}   &-10.30 &-10.57 & ~$-$10.17$\pm$0.13 &  2 &                  &    &~~$-$9.87~ \\                         
\ion{La}{ii}   &-10.32:&       & ~$-$10.21$\pm$0.17 &  4 &                  &    &~$-$10.91~ \\                         
\ion{Ce}{ii}   & -9.40 & -9.72 & ~~$-$9.47$\pm$0.18 & 11 &                  &    &~$-$10.46~ \\                         
\ion{Ce}{iii}  &       &       & ~~$-$7.63:         &  1 &                  &    &~$-$10.46~ \\                         
\ion{Pr}{ii}   &-10.40 &       & ~$-$10.16:         &  1 &                  &    &~$-$11.33~ \\                         
\ion{Pr}{iii}  &       &       & ~~$-$8.49$\pm$0.23 & 11 &                  &    &~$-$11.33~ \\                          
\ion{Nd}{ii}   & -9.30 & -9.56 & ~~$-$9.50$\pm$0.19 & 19 &                  &    &~$-$10.59~ \\                         
\ion{Nd}{iii}  &       & -8.53 & ~~$-$7.83$\pm$0.26 & 22 &                  &    &~$-$10.59~ \\                          
\ion{Sm}{ii}   & -9.50 &       & ~~$-$9.66$\pm$0.24 & 21 &                  &    &~$-$11.03~ \\                         
\ion{Eu}{ii}   & -9.40 & -9.94 & ~~$-$9.67$\pm$0.07 &  5 &~~$-$9.89$\pm$0.12&  4 &~$-$11.52~ \\                       
\ion{Gd}{ii}   & -9.45 &       & ~~$-$9.45$\pm$0.26 &  5 &                  &    &~$-$10.93~ \\                         
\ion{Tb}{ii}   &       &       & ~$-$10.2~$\pm$0.3: &  2 &                  &    &~$-$11.76~ \\                         
\ion{Tb}{iii}  &       &       & ~~$-$8.36$\pm$0.28 &  6 &                  &    &~$-$11.76~ \\                          
\ion{Dy}{ii}   &-10.00 &       & ~~$-$9.54$\pm$0.27 &  4 &                  &    &~$-$10.90~ \\                         
\ion{Dy}{iii}  &       &       & ~~$-$7.28$\pm$0.33 &  6 &                  &    &~$-$10.90~ \\                          
\ion{Ho}{ii}   &       &       & ~~$-$9.20:         &  1 &~$-$10.5~$\pm$0.2~&  4 &~$-$11.53~ \\                       
\ion{Er}{ii}   &       &       & ~~$-$9.90$\pm$0.29 &  5 &                  &    &~$-$11.11~ \\                       
\ion{Er}{iii}  &       &       & ~~$-$7.71$\pm$0.11 &  2 &                  &    &~$-$11.11~ \\                        
\ion{Tm}{iii}  &       &       & ~~$-$7.85:         &  1 &                  &    &~$-$12.04~ \\                        
\hline											     %
$\teff$ (K)    & $7900\pm200$   & $7420\pm170$    &\multicolumn{4}{c|}{$7500\pm130$}      & 5777  \\                            
$\logg$        & $4.20\pm0.15$  & $4.09\pm0.08$  &\multicolumn{4}{c|}{$4.10\pm0.15$}      & 4.44  \\                            
\vt\ (\kms)    & $1.50\pm0.15$  & $1.60\pm0.15$  &\multicolumn{4}{c|}{$1.20\pm0.20$}      &       \\                            
\hline											  
\end{tabular}
\end{center}
\end{footnotesize}
\end{table*}

We started our analysis calculating model atmospheres for \hd\ with the homogeneously distributed 
abundances taken from \citetalias{Bruntt08} and supplemented where necessary by the abundances from \citet{KRW96}.
Then the iterative procedure of stratification analysis was performed following the scheme outlined in
Sect.~\ref{sec:ss}. For each iteration we re-calculated atmospheric abundances of all available elements and
vertical stratification profiles of Si, Ca, Cr, and Fe. In total, four such iterations of the abundance
analysis and model atmosphere calculations were necessary to obtain a converged solution. At each iteration an
adjustment of $\teff$ by $\le$\,200~K was required to fit the observed spectral energy distribution and the hydrogen
line profiles. The final model atmosphere parameters inferred for \hd\ are $\teff$\,=\,7500~K and
$\logg$\,=\,4.1.

A comparison of the observed and calculated profiles of 17 \ion{Fe}{i} and 14 \ion{Fe}{ii} lines used for
stratification analysis is presented in Fig.~\ref{fig:fe}. These synthetic spectra were computed with the
vertical abundance distribution and atmospheric model obtained at the last iteration. Similar plots for Si, Ca,
and Cr are provided in Figs.~\ref{fig:si}--\ref{fig:cr} (Online material).

The variation of the stratification profiles of chemical elements during the iterative process is illustrated
in Fig.~\ref{fig:strat}. Neither the number of spectral lines used for the stratification analysis  with
\ddafit\ nor their atomic parameters were modified during the iteration procedure and, thus, all the changes in the
stratification profiles from one iteration to another are entirely due to changes of the model structure
and a small adjustment of $\teff$.
In general, the difference between the shape of the stratification profiles obtained on the first and last
iterations is relatively small for all considered elements. The positions of the abundance jumps did not change
much, although their amplitudes increased noticeably for Si and Ca. On the other hand, a large change of the
stratification parameters occurred at the second iteration, after stratified abundances were introduced in the
model atmosphere calculation for the first time. For this iteration we also had to adopt a 50~K lower effective
temperature compared to $\teff$\,=\,7500~K, which provided the best description of hydrogen lines and spectral
energy distribution for other iterations.

For all four elements we obtained a familiar picture of a step-like decrease of abundance with height. 
A slight distortion of abundance profiles in deeper layers apparent in Fig.~\ref{fig:strat} is caused by
a non-linear mapping from the column mass scale employed for vertical inversion to the standard optical 
depth used in the figure. For Cr and
Fe the transition region is narrow and is located between $\log\tau_{5000}$\,=\,$-0.5$ and 0.0. It is centered at
$\log\tau_{5000}$\,$\approx$\,$-1$ and is more extended for Si and Ca. All studied elements show a large
overabundance with respect to the sun in deep layers and significantly (Si, Ca) or marginally (Cr, Fe) sub-solar
element concentrations in the upper atmosphere. The formal error of the stratification parameters ranges from
0.4--0.8 dex for Ca and Si to 0.05--0.15~dex for Fe and Cr, with the upper atmospheric abundance  being more
uncertain. At the same time, position of abundance jump was determined with the precision of better than
0.1~dex in the $\log\tau_{5000}$ scale for all four elements.

Table\,\ref{tab:abund} summarizes the final mean abundances derived after the fourth iteration of the model
atmosphere calculations. Element concentrations are reported for 52 ions of 35 elements. For completeness we
also included mean abundances of the stratified elements. Our abundance estimates are compared with those
reported by \citet{KRW96} and \citetalias{Bruntt08} and with the solar abundances compiled by \citet{asplund}. We
find that \hd\ has a moderate CNO deficiency, almost solar Ca to Fe abundances  and an overabundance of Cr by
$\approx$\,0.8 dex.  The same overabundance is obtained for Co when hfs effects are taken into account. Ni
exhibits a 0.8~dex deficiency. In general, our mean abundance results agree well with those by \citetalias{Bruntt08}
except for C, S, Fe, \ion{Y}{ii}, Zr, and Ba. For the latter the difference in the adopted microturbulence 
parameter might be the reason for the discrepancy. \citetalias{Bruntt08} determined abundances for three out of $14$ REE
elements with stable isotopes, while we succeeded in estimating abundances for 12 REE species. For 6 REEs (Ce, Pr, Nd,
Tb, Dy, Er) abundances were obtained from the lines of the first and second ions separately.  Although \hd\ has
a moderate REE enhancement of 0.7--1.6 dex according to the results obtained using the first ions, the REE anomaly typical for roAp stars
\citep{RNW04} is clearly observed for all 6 REEs where abundances were obtained from the lines of two ionization
stages. This anomaly is $\approx$\,1.7--1.8 dex for Ce, Pr, Nd, and Tb and increases to 2.3 dex for Dy
and Er. This ionization imbalance is an indication of the accumulation of the REEs in the upper atmospheric
layers \citep{nd-NLTE,pr-NLTE}. Its effect on the model atmosphere calculations was investigated in detail by
\citet{hd24712}.

The fourth column in Table\,\ref{tab:abund}  gives abundances and standard deviations obtained  with the
\width\ code from $n$ spectral lines neglecting hfs effects, while in the next column one finds abundances
calculated with \synthmag\ taking hfs effects into account.  The largest hfs effect is evident for \ion{Ho}{ii}
lines, where the final abundance decreases by about an order of magnitude. The hfs effect on \ion{Eu}{ii} lines
is the smallest one and does not exceed $0.25$\,dex.  The detailed treatment of the hyperfine splitting of the
\ion{Co}{i} lines increases both the discrepancy with \ion{Co}{ii} and the standard deviation  compared to the
\width\ analysis. This could be a signature of Co stratification in the atmosphere of \hd. Separate hfs
analysis of the three moderately strong lines in the 3500--3600~\AA\ region  (\eqw\,=\,40--82~m\AA) and five
weak to medium strength lines in 5300--5400~\AA\ region (\eqw\,=\,13--42~m\AA) gives $\log (Co/N_{\rm
tot})=-6.77\pm0.16$ and $\log (Co/N_{\rm tot})=-6.10\pm0.09$, respectively. This indicates possible Co
accumulation in deep layers of the stellar atmosphere. The absence of hfs constants for \ion{Co}{ii} lines does
not allow us to carry out a formal Co stratification analysis. In addition, our abundance analysis shows that
Sr might be strongly stratified in the atmosphere  of \hd\ similar to other Ap stars like, for example,
HD~133792 \citep{vip}. However, the lack of significant number of suitable  unblended lines again prevented
detailed stratification analysis for this element.


\subsection{Spectral energy distribution and photometry}

\begin{figure*}
\includegraphics[height=\hsize,angle=90]{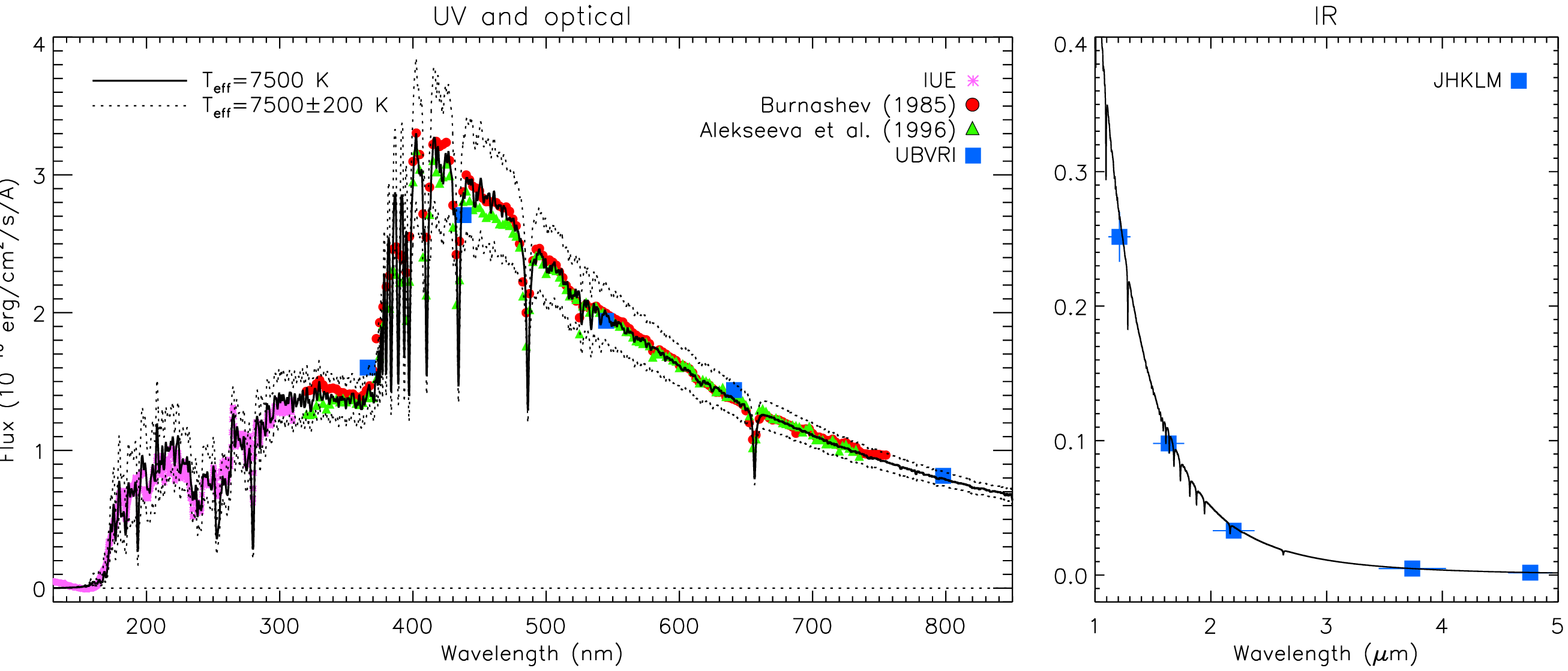}
\caption{Comparison between the observed (\textit{symbols}) and theoretical (\textit{lines}) 
spectral energy distribution of \hd. Photometric and spectrophotometric observations, 
covering the wavelength range from the ultraviolet and
optical (\textit{left panel}) to the infrared (\textit{right panel}), are discussed in the
text (see Sect.~\ref{sec:obs}). The solid line shows theoretical energy distribution computed 
for the final self-consistent model atmosphere of \hd\ with parameters $\teff$\,=\,7500~K and 
$\logg$\,=\,4.1. The dotted lines illustrate the effect of changing $\teff$ by $\pm200$~K.
The \llm\ flux spectra are convolved with a $FWHM=10$~\AA\ Gaussian profile.}
\label{fig:sed}
\end{figure*}

\onlfig{7}{
\begin{figure*}
\centerline{\includegraphics[height=15cm,angle=90]{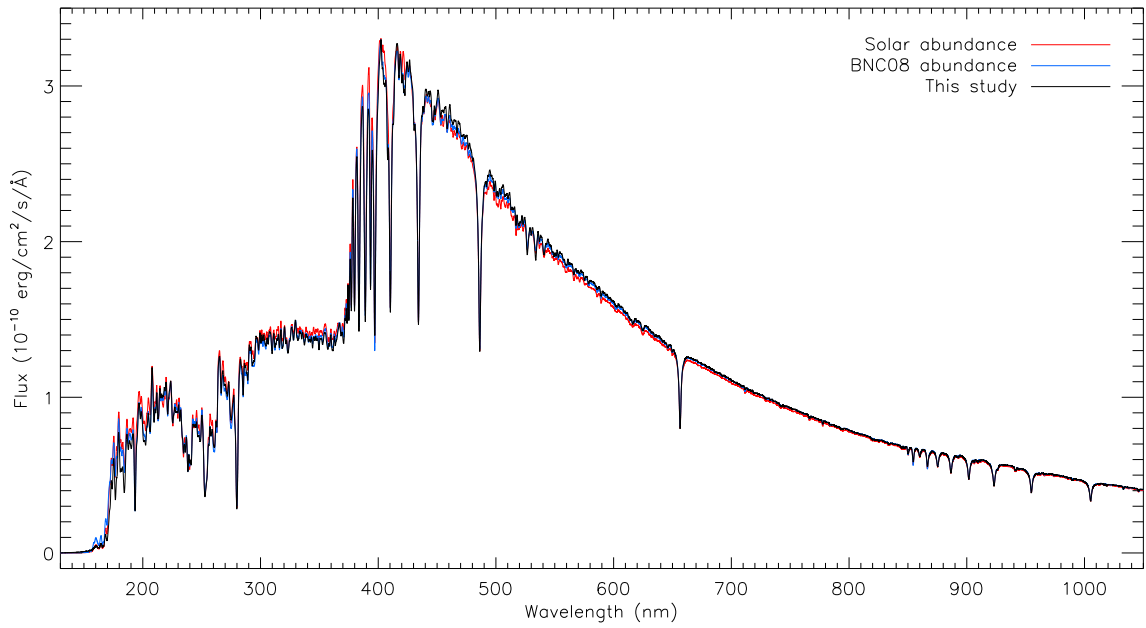}}
\caption{Theoretical spectral energy distributions for the model atmosphere parameters
$\teff$\,=\,7500~K, $\logg$\,=\,4.1 and different chemical compositions of \hd. The \llm\ flux spectra, convolved
with a $FWHM=10$~\AA\ Gaussian profile, are presented for our final self-consistent 
model atmosphere with stratification included (\textit{black line}), chemically-homogeneous
solar abundance model (\textit{red line}), and a model computed for \citetalias{Bruntt08} abundances 
(\textit{blue line}).}
\label{fig:sedt}
\end{figure*}
}

\begin{figure*}
\sidecaption
\includegraphics[width=12cm]{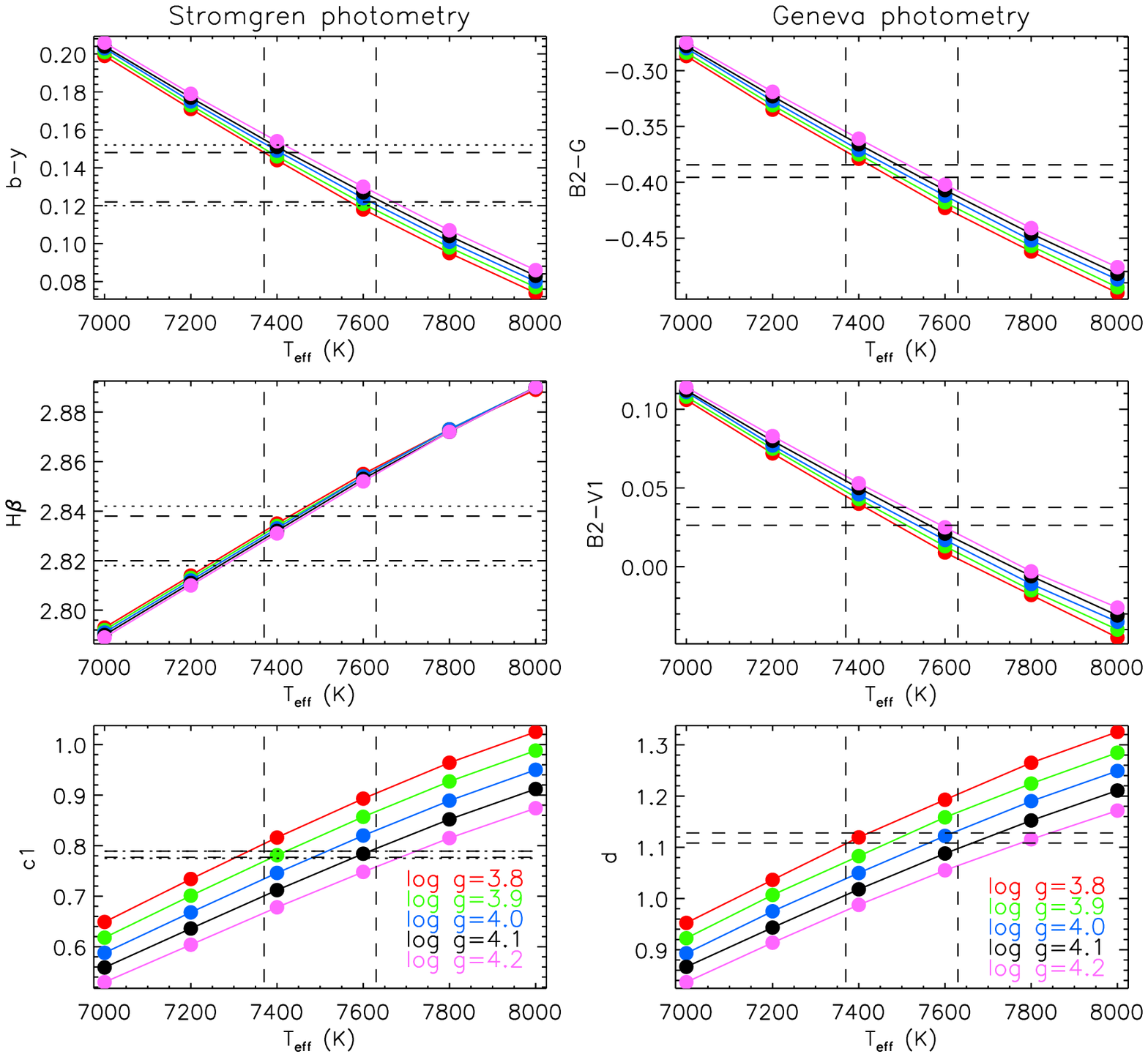}
\caption{Dependence of the synthetic Str\"omgren (\textit{left panel}) and Geneva
(\textit{right panel}) photometric parameters on the $\teff$ and $\logg$ for the models with
stratified chemical composition. The theoretical temperature ($b-y$, H$\beta$
for the Str\"omgren and $B2-G$, $B2-V1$ for the Geneva photometry) and gravity ($c_1$
for the Str\"omgren and $d$ for the Geneva photometry) indicators are compared with
the corresponding observed parameters. Those are indicated by the horizontal dashed lines 
showing 
$\pm1\sigma$ ranges and by the dash-dotted lines (Str\"omgren photometry only) showing
the minimum and maximum observed values. The vertical dashed lines correspond to the 
$7500\pm130$~K $\teff$ range. 
}
\label{fig:colors}
\end{figure*}

We used the observed spectral energy distribution of \hd\ as the main effective temperature
indicator. Previous attempts to model spectrophotometry of magnetic Ap stars
\citep{AR00,LS08} could utilize only the information about the shape of the energy
distribution because observations were often given in relative flux units. Furthermore, the stellar  angular
diameters were generally unknown for Ap stars and had to be treated as free parameters. 
For \hd\ we are in the unique position to obtain a very precise estimate of the model atmosphere parameters from the
observed stellar flux thanks to the
recent direct, interferometric
determination of the angular diameter, $\theta=1.105\pm0.037$~mas \citepalias{Bruntt08}.
Fig.~\ref{fig:sed} presents a comparison of the photometric and
spectrophotometric measurements of \hd, discussed in detail in Sect.~\ref{sec:obs}, with the
theoretical flux distribution computed for the models with individual abundance and
stratification included. We find that observations are best fitted by the model with
$\teff$\,=\,7500~K. The uncertainty of this effective temperature estimate, related to
fitting ambiguities and discrepancy between the two spectrophotometric scans available for
\hd, does not exceed 50~K. At the same time, a $\teff$ error due to the angular diameter uncertainty is
estimated to be 120~K. Thus, our analysis of the spectral energy distribution yields an
effective temperature with the overall accuracy of about 130~K.

Following \citetalias{Bruntt08}, we adopted $\logg$\,=\,4.1 for the modeling of the observed
flux of \hd. Changes of the surface gravity by 0.1--0.2~dex lead to a negligible modification
of the theoretical spectral energy distribution and do not alter results of the $\teff$
determination.

In the case of \hd, spectrophotometry is not very sensitive to the adopted atmospheric
chemical composition and stratification. In Fig.~\ref{fig:sedt} (Online material) we compare
\llm\ fluxes calculated for our self-consistent, chemically-stratified model atmosphere and
for the models adopting homogeneous solar abundance and the chemical composition published
by \citetalias{Bruntt08}. All three models predict similar flux distributions, although
calculations with the non-solar abundances exhibit a more substantial redistribution of flux
from the UV region to the optical and IR. The chemically-stratified model predicts a
marginally higher flux for $\lambda\le200$~nm compared to the model with homogeneous
individual abundances.

Observations in the Str\"omgren and Geneva photometric systems are often used for 
determination of the atmospheric parameters of magnetic CP stars \citep{HN93,netopil}.
Therefore, we examined an agreement of the photometric parameters of \hd\ predicted by our
self-consistent stratification models with the observed photometric properties of this star.
For stars with a temperature close to that of \hd, the $b-y$ and H$\beta$ parameters of the
Str\"omgren photometry can be used as $\teff$ indicators, while $c_1$ is mainly sensitive to the
surface gravity \citep{MD85,N93}. For the Geneva photometry, the effective temperature 
calibrations are usually based upon
the $B2-G$ and $B2-V1$ indices and gravity can be determined with
the help of the reddening-free $d$-parameter \citep{HN93,K97}. Fig.~\ref{fig:colors} presents
these $\teff$ and $\logg$ indicators for the entire model atmosphere grid computed with the
final abundances and chemical stratification. The observed range of the Str\"omgren
photometric observations of \hd\ is indicated by the horizontal lines. Uncertainties of the
Geneva photometric parameters are calculated taking into account an observational error of
0.004~mag \citep{geneva}.

\begin{figure}
\includegraphics[width=\hsize]{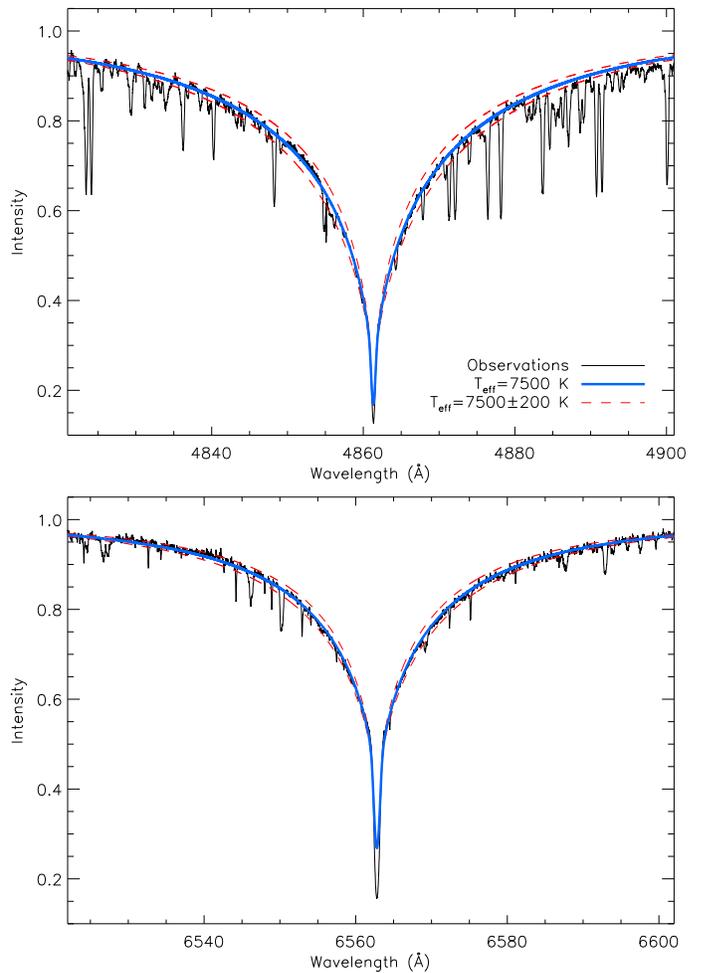}
\caption{
Comparison of the observed and theoretical H$\alpha$ and H$\beta$ line profiles of \hd.
Observations are shown by the thin line. Thick line gives the hydrogen line profiles computed for our
final self-consistent, chemically-stratified 
model atmosphere with $\teff$\,=\,7500~K, $\logg$\,=\,4.1. Dashed lines show the effect
of changing $\teff$ by $\pm200$~K.
}
\label{fig:hlines}
\end{figure}

\onlfig{10}{
\begin{figure*}
\centerline{\includegraphics[width=9cm]{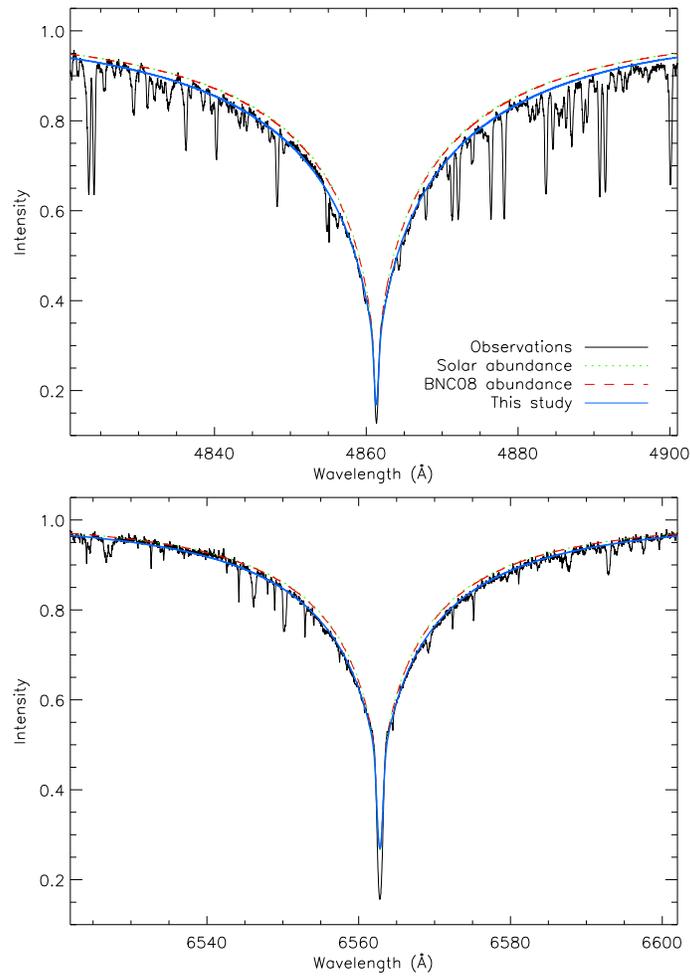}}
\caption{Comparison of the observed and theoretical H$\alpha$ and H$\beta$ line profiles of \hd. 
Observations are shown by the black line. Theoretical calculations with parameters $\teff$\,=\,7500~K, 
$\logg$\,=\,4.1 are presented for our final self-consistent 
model atmosphere with stratification included (\textit{blue line}), chemically-homogeneous
solar abundance model (\textit{green line}), and a model computed for \citetalias{Bruntt08} abundances 
(\textit{red line}).}
\label{fig:hlinest}
\end{figure*}
}

\begin{figure}[!th]
\includegraphics[width=\hsize]{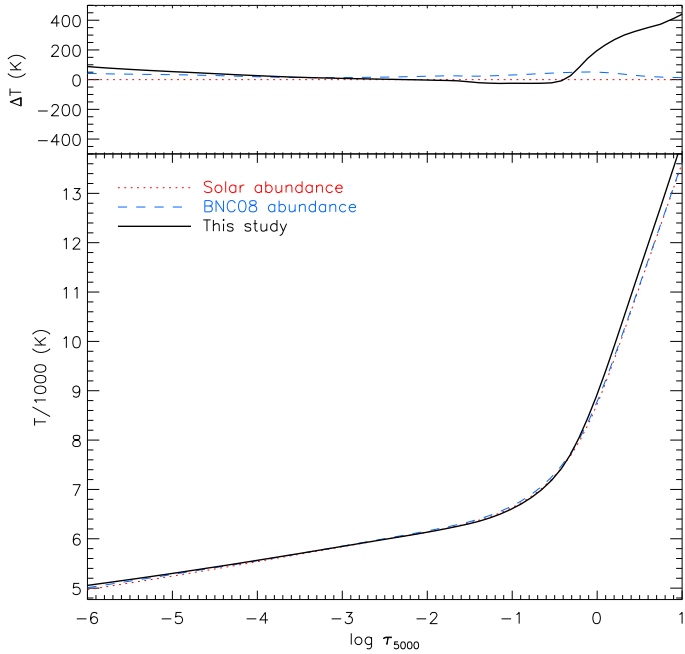}
\caption{Temperature as a function of standard optical depth, $\tau_{5000}$, for the final self-consistent 
model atmosphere of \hd\ with $\teff$\,=\,7500~K, $\logg$\,=\,4.1 (\textit{solid line}) compared to the 
temperature structure of the models with the same atmospheric parameters but using homogeneous solar (\textit{dotted line})
and \citetalias{Bruntt08} (\textit{dashed line}) abundances. The upper panel shows the 
temperature differences with
respect to the solar-abundance case.}
\label{fig:atm}
\end{figure}

Generally, Fig.~\ref{fig:colors} shows an excellent agreement between the observed and 
synthetic $b-y$, $B2-G$, and $B2-V1$ indices for $\teff$\,=\,$7500\pm130$~K estimated from
the spectrophotometry. The H$\beta$ parameter indicates a somewhat lower $\teff$
on average, but it is still consistent with our model predictions if the scatter of photometric
measurements 
is taken into account. A direct spectroscopic comparison between the observed and computed
hydrogen line profiles presented below provides a more robust method to estimate $\teff$ from
the Balmer lines. The observed $c_1$ index suggests $\logg$\,=\,$4.0\pm0.15$ while
$\logg$\,=\,$3.9\pm0.15$ fits the observed Geneva $d$-parameter. Both estimates do not
contradict $\logg$\,=\,$4.08\pm0.09$ determined by \citetalias{Bruntt08} \txtbf{with} stellar
evolutionary models.

\subsection{Hydrogen lines}

Due to its relatively low temperature, the hydrogen line profiles of \hd\ are mostly
sensitive to $\teff$. Thus, we have used the \halpha\ and \hbeta\ lines to verify the effective
temperature of \hd\ determined from the stellar flux distribution. The comparison between the
observed hydrogen lines and the \synthmag\ calculations for the final chemically-stratified
models is illustrated in Fig.~\ref{fig:hlines}. An effective temperature of $7470\pm100$~K
yields the best fit to the \halpha\ and \hbeta\ profiles. The error takes into account
fitting uncertainty and  $\approx$\,1\% ambiguity of the continuum normalization in the regions
around broad hydrogen line wings. 

Unlike the spectral energy distribution, hydrogen lines are significantly modified when
chemical stratification is included in the model atmosphere calculations.
Fig.~\ref{fig:hlinest} (Online material) shows that theoretical Balmer line profiles  become
systematically weaker for the models which neglect chemical stratification. Consequently, the
stellar effective temperature will be overestimated by 200~K if such models are to be used
for fitting the hydrogen line wings. This interesting result is explained in
Fig.~\ref{fig:atm}, which shows temperature as a function of continuum optical depth for models with
the same atmospheric parameters but different chemical compositions. The calculations with
the homogeneous individual abundance do not deviate strongly from the solar-composition
model. On the other hand, the presence of a large Cr, Si, and Fe overabundance in deep
photospheric layers of the chemically-stratified model leads to a significant backwarming
effect, corresponding to a temperature increase by 200--400~K for $\log\tau_{5000}\ge0$. The
strengthening of the hydrogen line wings reflects this change in the model atmosphere
structure.

\section{Conclusions and discussion}
\label{sec:concl}

An empirical investigation of the vertical distribution of chemical elements in the
atmospheres  of CP stars is of great importance for testing predictions of the microscopic
particle diffusion theory. Such investigations yield stratification profiles for
different chemical elements that could be directly confronted with the diffusion calculations.

In this paper we determined empirical vertical distributions of Si, Ca, Cr, and Fe in the
atmosphere of the brightest rapidly oscillating Ap star \hd. Stratification analysis was
complemented with the most detailed study of average abundances ever conducted for \hd.
Unlike many previous empirical stratification studies of magnetic CP stars, we made an effort
to attain a self-consistency between the inferred chemical properties of the stellar
atmosphere and its temperature-pressure structure. This was achieved by using an iterative
procedure of deriving stratification and mean abundances from the observed high-resolution
spectra and subsequent recomputing the model atmosphere structure.

\txtbf{Despite a significant improvement of this stellar atmosphere modeling methodology
compared to previous studies of cool Ap stars, the analysis presented in this paper has
several limitations, stemming from its empirical character. The lines available in the
spectra of \hd\ can be used to establish stratification of only four chemical elements and
do not allow exploring stratification outside the optical depth range of approximately
$\log\tau_{5000}=-3$ to $+0.5$. Furthermore, observations do not allow resolving possible variation of
chemical stratification profiles over the stellar surface, which might be present due to
magnetic field effects. Potential NLTE effects are also neglected.}

The availability of the photometric measurements in different systems, spectrophotometric
scans, hydrogen line profiles and, uniquely for \hd, direct interferometric measurement of
the angular diameter allowed us to derive accurate atmospheric parameters and test the
influence of non-solar, inhomogeneous abundances on fitting different types of observations.

The final atmospheric parameters of \hd\ were found to be $\teff=7500\pm130$\,K and
$\logg=4.10\pm0.15$. Together with the chemical stratification and mean abundances derived in
our study, these parameters reproduce all photometric, spectrophotometric, and spectroscopic
constraints available for \hd. Our atmospheric parameters are in excellent agreement with 
$\teff=7420\pm170$~K, $\logg=4.09\pm0.08$ determined by \citetalias{Bruntt08} from the
angular diameter measurement, bolometric flux and stellar structure calculations. 
This \txtbf{is an indication} that modern model atmosphere
analyses of cool Ap stars are trustworthy and are not affected by significant systematic
biases.

The vertical distribution of chemical elements in the atmosphere of \hd\ was found to be
similar to the abundance profiles inferred for other cool Ap stars. All four elements
analysed for \hd\ exhibit a large overabundance in the lower atmospheric layers and a solar
or subsolar abundance above $\log\tau_{5000}\approx-1$. We found that the stratification
profiles of chemical elements derived using a model atmosphere computed with homogeneous
abundances could slightly differ from those obtained using a self-consistent analysis.
However, in general, these changes do not alter significantly the overall energy balance in
the stellar atmosphere.

For the atmospheric parameters of \hd, chemical stratification has a minor
influence on the spectral energy distribution in comparison to chemically-homogeneous models.
Thus, interpretation of the spectrophotometric scans will not be significantly in error if
stratification is ignored. Similar result was obtained in our study of another cool Ap star
HD\,24712 \citep[$\teff=7250$\,K, $\logg=4.1$][]{hd24712}. We caution, however, that the
validity of this finding should not be extended outside the $\teff$\,=\,7000--8000~K range
addressed in our investigations. \txtbf{Hotter Ap stars might have a different abundance
stratification that makes a larger impact on the atmospheric structure and 
stellar parameter determination.}

On the other hand, we found that the backwarming effect introduced by concentration of
iron-peak elements in the lower atmospheric layers of the chemically-stratified model leads
to significant strengthening of the hydrogen line wings. The corresponding reduction of
$\teff$ inferred from the Balmer lines amounts to $\approx$\,200~K. In the case of \hd, only
chemically-stratified models are able to reproduce the \halpha\ and \hbeta\ profiles with a
$\teff$ consistent with the photometric and spectrophotometric observations.

The effective temperature determined in our study is significantly lower \txtbf{than} $\teff=7900\pm200$~K
found in the previous spectroscopic analysis of \hd\ \citep{KRW96}. The origin of this discrepancy
is twofold. On the one hand, \citet{KRW96} analysed the star using standard model atmospheres, which
neglected a non-solar chemical composition and element stratification. This led to an overestimation
of $\teff$ obtained with the help of photometric techniques and hydrogen line wing fitting. On the
other hand, the final $\teff$ was obtained by \citet{KRW96} by reducing abundance scatter for a set
of \ion{Fe}{i} and \ion{Fe}{ii} lines, assuming a homogeneous distribution of this element with
height which seemed to be a reasonable approximation at that time. We now know that this assumption
is incorrect and that, due to their different formation heights, the Fe lines of different
excitation potentials and ionization degree give different apparent abundance  due to stratification
effects in cool Ap stars. In this situation the criterium of minimum abundance scatter does not 
allow one to recover the true stellar parameters.

Using the fundamental $\teff$ and a luminosity of $L=10.51\pm0.60 L_{\odot}$ estimated from the
Hipparcos parallax, \citetalias{Bruntt08} determined a stellar mass, $M$\,$\approx$\,1.7 $M_{\odot}$, with
the help of stellar structure models. We confirmed this result using Padova evolutionary tracks
\citep{padova}. Adopting our spectroscopic $\teff$ and the same luminosity as in
\citetalias{Bruntt08}, we found $M=1.69\pm0.02 M_{\odot}$ for the $Z=0.019$ grids. The small formal
uncertainty of the stellar mass is unrealistic given the lack of knowledge about the overall metal
content of \hd. For example, assuming $Z=0.008$, which is closer to the recent determination of the
solar atmospheric metallicity \citep{asplund}, yields a mass of $M=1.52\pm0.02 M_{\odot}$. For both
metallicities the stellar radius predicted by the Padova models is in excellent agreement with the
value of $R=1.967\pm0.066 R_{\odot}$ determined by \citetalias{Bruntt08}. The $Z=0.008$--0.019
metallicity interval represents a reasonable estimate of the range of the bulk metal content of
nearby field stars \citep{clust}. Unknown initial helium abundance and variation in the treatment of
convection in stellar structure calculations may introduce additional uncertainties in the mass
determination for late-A stars \citep{cunha}. Thus, despite availability of precise estimates of
the atmospheric parameters, radius and luminosity, we can only conclude that the mass of \hd\ is
likely to be in the range from about 1.5\,$M_{\odot}$ to 1.7\,$M_{\odot}$.

In the light of new, reliable determination of the atmospheric and fundamental parameters of \hd,
one can use this star to test photometric temperature calibrations of cool Ap stars. These
calibrations were recently summarized and improved by \citet{netopil}.  For magnetic Ap stars with
$\teff\le9000$~K they provided a new $B-V$ calibration and recommended using the $B2-G$ calibration
by \citet{gencal} and the $uvby\beta$ calibration by \citet{N93} for the Geneva and Str\"omgren
photometric systems, respectively. Application of these photometric temperature calibrations gives
$\teff$($B-V$)\,=\,7940~K, $\teff$(Geneva)\,=\,7678~K, and $\teff$(Str\"omgren)\,=\,7692~K. In
comparison, an earlier Str\"omgren photometry calibration by \citet{MD85} yields $\teff$\,=\,7955~K.
Thus, all calibrations overestimate effective temperature of \hd\ by 200--450~K while performing
much better for other cool Ap stars \citep{netopil}. An unusually high abundance of Fe in the 
atmosphere of \hd\ compared to other cool magnetic Ap stars with $\teff\le7500$~K \citep{TR05} might
be the reason for deviating photometric behaviour of this star. 

A large difference in the mean iron-peak element abundances of \hd\ and of HD\,24712, which we
analysed using a similar self-consistent model atmosphere technique \citep{hd24712}, explains the
different influence of stratified abundances on the model structure of the two stars. Both Fe and Cr
are significantly more abundant in \hd. Consequently, their stratification changes the atmospheric
structure to a much greater extent than for HD\,24712 inspite of a qualitatively similar
stratification profiles inferred for the two stars.

Some of the properties of \hd\ might be connected with the special aspect angle from which this star
is visible from Earth. The presence of narrow line profiles for the 4.46$^{\rm d}$ rotation period
of \hd\ indicates a relatively low inclination angle ($i\approx35\degr$). The restricted visible
part of the stellar surface might not be representative of the whole star if, for example, it is
covered by predominantly horizontal magnetic fields and by spots of enhanced iron-peak element
abundance. Precise spectroscopic and polarization observations of \hd\ over the entire rotation
cycle are required to investigate the surface magnetic and abundance geometry.

\begin{acknowledgements}
This work was supported by 
FWF Lisa Meitner grant Nr. M998-N16 (DS), FWF P17890-N2 (TR), RFBR 08-02-00469-a and
Presidium RAS Programme ``Origin and evolution of stars and galaxies'' to TR.
\end{acknowledgements}


\begin{thebibliography}{}
\bibitem[Adelman \& Rayle, 2000]{AR00}{Adelman}, S.~J. \& {Rayle}, K.~E. 2000, \aap, 355, 308
\bibitem[Alecian \& Stift, 2007]{diff_alecian}Alecian, G. \& Stift, M. J. 2007, \aap, 475, 659
\bibitem[Alekseeva et al., 1996]{alekseeva}Alekseeva, G.~A., Arkharov, A.~A., Galkin, V. D., et al. 1996, Baltic Astron., 5, 603
\bibitem[Asplund et al., 2005]{asplund}Asplund, M., Grevesse, N., \& Sauval, A.J. 2005, in \textit{Cosmic Abundances as Records of Stellar Evolution and Nucleosynthesis}, eds. T.G. Barnes III, F.N. Bash, ASP Conf. Ser., 336, 25
\bibitem[Babel, 1992]{babel}Babel, J. 1992, \aap, 263, 232
\bibitem[Bagnulo et al., 2001]{BWD01}Bagnulo, S., Wade, G. A., Donati, J.-F., et al. 2001, \aap, 369, 889
\bibitem[Bagnulo et al., 2006]{clust}{Bagnulo}, S., {Landstreet}, J.~D., {Mason}, E. et al. 2006, \aap, 450, 777
\bibitem[Ballester et al., 2000]{pipeline}Ballester, P., Modigliani, A., Bitquin, O., et al. 2000, Messenger, 101, 31
\bibitem[Bard \& Kock, 1994]{bard}Bard, A. \& Kock, M. 1994, \aap, 282, 1014 
\bibitem[Berry et al., 1971]{berry}Berry, H.G., Bromander, J., Curtis, L.J., \& Buchta, R. 1971, \physscr, 3, 125
\bibitem[Bessell et al., 1998]{bessell}{Bessell}, M.~S., {Castelli}, F., \& {Plez}, B. 1998, \aap, 333, 231
\bibitem[Bruntt, 2007]{wire}Bruntt, H. 2007, CoAst, 150, 326
\bibitem[Bruntt et al., 2008]{Bruntt08}Bruntt, H., North, J.R., Cunha, M. et~al. 2008, \mnras, 386, 2039
\bibitem[Burnashev, 1985]{burnashev}Burnashev, V.~I., 1985, Abastumanskaya Astrofiz. Obs., Byull., 59, 83
\bibitem[Cowley et al., 2001]{cowley_cwa}{Cowley}, C.~R., {Hubrig}, S., {Ryabchikova}, T.~A., et al. 2001, \aap, 367, 939
\bibitem[Cunha et al., 2003]{cunha}{Cunha}, M.~S., {Fernandes}, J.~M.~M.~B., \& {Monteiro}, M.~J.~P.~F.~G. 2003, \mnras, 343, 831
\bibitem[Dekker et al., 2000]{uves}Dekker, H., D'Odorico, S., Kayfer, A., Delabre, B., \& Kotzlowski, H. 2000, Proc. SPIE, 4008, 534
\bibitem[Engels et al., 1981]{engels}{Engels}, D., {Sherwood}, W.~A., {Wamsteker}, W., \& {Schultz}, G.~V. 1981, \aaps, 45, 5
\bibitem[Girardi et al., 2000]{padova}Girardi, L., Bressan, A., Bertelli, G., \& Chiosi, C. 2000, \aaps, 141, 371
\bibitem[Groote \& Kaufmann, 1983]{jhklm}{Groote}, D. \& {Kaufmann}, J.~P. 1983, \aaps, 53, 91
\bibitem[Hauck \& North, 1982]{gencal}{Hauck}, B. \& {North}, P. 1982, \aap, 114, 23
\bibitem[Hauck \& North, 1993]{HN93}{Hauck}, B. \& {North}, P. 1993, \aap, 269, 403
\bibitem[Holt et al., 1999]{HSR99}Holt, R.A., Scholl, T.J., \& Rosner, S.D. 1999, \mnras, 306, 107  
\bibitem[Hui-Bon-Hoa et al., 2000]{diff_hui}Hui-Bon-Hoa, A., LeBlanc, F., \& Hauschildt, P. H. 2000, \apj, 535, L43
\bibitem[Johnson et al., 1966]{ubvri}Johnson, H. L., Iriarte, B., Mitchell, R. I., \& Wisniewskj, W. Z. 1966, Comm. Lunar Plan. Lab., 4, 99
\bibitem[Khan \& Shulyak, 2006]{zeeman_paper2}Khan, S. \& Shulyak, D. 2006, \aap, 448, 1153
\bibitem[Kochukhov, 2007]{synthmag07}Kochukhov, O. 2007, in \textit{Physics of Magnetic Stars}, eds. D.O.~Kudryavtsev, I.I~Romanyuk, Nizhnij Arkhyz., 109
\bibitem[Kochukhov et al., 2002]{kochukhov_cwa}{Kochukhov}, O., {Bagnulo}, S., \& {Barklem}, P.~S. 2002, \apj, 578, L75
\bibitem[Kochukhov et al., 2004]{hr3831}{Kochukhov}, O., {Drake}, N.~A., {Piskunov}, N., \& {de la Reza}, R. 2004, \aap, 424, 935
\bibitem[Kochukhov et al., 2005]{zeeman_paper1}{Kochukhov}, O., {Khan}, S., \& {Shulyak}, D. 2005, \aap, 433, 671
\bibitem[Kochukhov et al., 2006]{vip}Kochukhov, O., Tsymbal, V., Ryabchikova, T., Makaganyk, V., \& Bagnulo, S. 2006, \aap, 460, 831
\bibitem[Kochukhov et al., 2007]{KRW07}{Kochukhov}, O., {Ryabchikova}, T., {Weiss}, W.~W., et al. 2007, \mnras, 376, 651
\bibitem[Kochukhov \& Ryabchikova, 2001]{KR01}Kochukhov, O. \& Ryabchikova, T. 2001, \aap, 377, L22
\bibitem[Kunzli et al., 1997]{K97}{Kunzli}, M., {North}, P., {Kurucz}, R.~L., \& {Nicolet}, B. 1997, \aaps, 122, 51  
\bibitem[Kupka et al., 1996]{KRW96}Kupka, F., Ryabchikova, T.A., Weiss, W.W., et al. 1996, \aap, 308, 886
\bibitem[Kupka et al., 1999]{vald2}Kupka, F., Piskunov, N., Ryabchikova, T. A., Stempels, H. C., \& Weiss, W. W. 1999, \aaps, 138, 119
\bibitem[Kurtz et al., 1994]{kurtz}{Kurtz}, D.~W., {Sullivan}, D.~J., {Martinez}, P., \& {Tripe}, P. 1994, \mnras, 270, 674
\bibitem[Kurucz, 2008]{ATOMS}Kurucz, R.L. 2008, http://cfaku5.cfa.harvard.edu/ATOMS
\bibitem[Kurucz, 1993]{a9}Kurucz, R.L. 1993, Kurucz CD-ROM 13, Cambridge, SAO
\bibitem[Lawler et al., 2001]{Eu2}Lawler, J.E., Wickliffe, M.E., Den Hartog, E.A., \& Sneden, C. 2001, \apj, 563, 1075  
\bibitem[Lawler et al., 2004]{Ho2}Lawler, J.E., Sneden, C., \& Cowan, J.J. 2004, \apj, 604, 850  
\bibitem[LeBlanc \& Monin, 2004]{LM04}LeBlanc, F. \& Monin, D. 2004, in {\it IAU Symp. 224, The A-Star Puzzle}, eds. J. Zverko, J. Ziznovsky, S.J. Adelman, and W.W. Weiss, Cambridge University Press, 193
\bibitem[\txtbf{LeBlanc et al.}, 2009]{LMH09}LeBlanc, F., Monin, D., Hui-Bon-Hoa, A., \& Hauschildt, P. H. 2009, \aap, in press 
\bibitem[Lipski \& St{\c e}pie{\'n}, 2008]{LS08}{Lipski}, {\L}. \& {St{\c e}pie{\'n}}, K. 2008, \mnras, 385, 481
\bibitem[Martin et al., 1988]{Martin}Martin, G.A., Fuhr, J.R., \& Wiese, W.L. 1988, J. Phys. Chem. Ref. Data, 17, Suppl.3
\bibitem[Mashonkina et al., 2005]{nd-NLTE}Mashonkina, L., Ryabchikova, T., \& Ryabtsev, A. 2005, \aap, 441, 309
\bibitem[Mashonkina et al., 2008]{pr-NLTE}Mashonkina, L., Ryabchikova, T., Ryabtsev, A., \& Kildiyarova, R. 2008, \aap, 495, 297
\bibitem[Michaud, 1970]{michaud}Michaud, G. 1970, \apj, 160, 641
\bibitem[Moon \& Dworetsky, 1985]{MD85}{Moon}, T.~T. \& {Dworetsky}, M.~M. 1985, \mnras, 217, 305
\bibitem[Napiwotzki et al., 1993]{N93}{Napiwotzki}, R., {Schoenberner}, D., \& {Wenske}, V. 1993, \aap, 268, 653
\bibitem[Netopil et al., 2008]{netopil}{Netopil}, M., {Paunzen}, E., {Maitzen}, H.~M., {North}, P., \& {Hubrig}, S. 2008, \aap, 491, 545
\bibitem[O'Brian et al., 1991]{obrian}O'Brian, T.R., Wicklife, M.E., Lawler, J.E., Whaling, W., \& Brault, J.W. 1991, JOSA, B8, 1185
\bibitem[Pickering, 1996]{hfs-Co1}Pickering, J.C. 1996, \apjs, 107, 811
\bibitem[Piskunov et al., 1995]{vald1}Piskunov, N. E., Kupka, F., Ryabchikova, T. A., Weiss, W. W., \& Jeffery, C. S. 1995, \aaps, 112, 525
\bibitem[Raassen \& Uylings, 1998]{Raassen1998}Raassen, A.J.J. \& Uylings, P.H.M. 1998, \aap, 340, 300
\bibitem[Ralchenko et al., 2008]{NIST}Ralchenko, Yu., Kramida, A.E., Reader, J., and NIST ASD Team 2008,
NIST Atomic Spectra Database (version 3.1.5), National Institute of Standards and Technology, Gaithersburg, MD.
\bibitem[Ryabchikova, 2005]{TR05}Ryabchikova, T. 2005, Astron. Letters, 31, 388 
\bibitem[Ryabchikova, 2008]{TR08}Ryabchikova, T. 2008, CoSka, 38, 257 
\bibitem[Ryabchikova et al., 2002]{RPK02}Ryabchikova, T., Piskunov, N., Kochukhov, O. et al. 2002, \aap, 384, 545
\bibitem[Ryabchikova et al., 2004]{RNW04}Ryabchikova, T., Nesvacil, N., Weiss, W.W., et al. 2004, \aap, 423, 705
\bibitem[Ryabchikova et al., 2005]{RLK05}Ryabchikova, T., Leone, F., \& Kochukhov, O. 2005, \aap, 438, 973
\bibitem[Ryabchikova et al., 2007]{RSK07}{Ryabchikova}, T., {Sachkov}, M., {Kochukhov}, O., \& {Lyashko}, D. 2007, \aap, 473, 907
\bibitem[Ryabchikova et al., 2008]{RKB08}Ryabchikova, T., Kochukhov, O., \& Bagnulo S. 2008, \aap, 480, 811
\bibitem[Rufener, 1988]{geneva}Rufener F. 1988, \textit{Catalogue of Stars Measured in the Geneva Observatory Photometric System}, Geneva Observatory
\bibitem[Shulyak et al., 2004]{llm}Shulyak, D., Tsymbal, V., Ryabchikova, T., St\"utz\, Ch., \& Weiss, W. W. 2004, \aap, 428, 993
\bibitem[Shulyak et al., 2008]{hd137509}Shulyak, D., Kochukhov, O., \& Khan, S. 2008, \aap, 487, 689
\bibitem[Shulyak et al., 2009]{hd24712}Shulyak, D., Ryabchikova, T., Mashonkina, L., \& Kochukhov, O. 2009, \aap, submitted
\bibitem[Schulz-Gulde, 1969]{Schulz}Schulz-Gulde, E. 1969, \jqsrt, 9, 13
\bibitem[Wilke, 2003]{Wilke03}Wilke, R. 2003, PhD thesis, Heinrich-Heine-Universit\"at, D\"usseldorf
\end{thebibliography}
\end{document}